\documentclass[prd]{revtex4}
\usepackage[dvips]{graphicx}

\begin{document}

\title{Nonlinear Evolution of Cosmic Magnetic Fields and 
Cosmic Microwave Background Anisotropies}

\author{Hiroyuki Tashiro}
\affiliation{Department of Physics, Kyoto University, Kyoto 606-8502, Japan
}
\email{htashiro@tap.scphys.kyoto-u.ac.jp}
\author{Naoshi Sugiyama}
\affiliation{Division of Theoretical Astronomy, National Astronomical Observatory,
 Japan, Mitaka, Tokyo 181-8588 Japan }
\author{Robi Banerjee}
\affiliation{Department of Physics and Astronomy, McMaster University,
  Hamilton, Ontario L8S 4M1, Canada
}
\date{\today}


\begin{abstract}

In this work we investigate the effects of primordial magnetic fields
on cosmic microwave background anisotropies (CMB). Based on
cosmological magneto-hydro dynamic (MHD) simulations~\cite{b-j-paper}
we calculate the CMB anisotropy spectra and polarization induced by
fluid fluctuations (Alfv\'en modes) generated by primordial magnetic
fields. The strongest effect on the CMB spectra comes from the
transition epoch from a turbulent regime to a viscous regime.  The
balance between magnetic and kinetic energy until the onset of the
viscous regime provides a one to one relation between the comoving
coherence length $L$ and the comoving magnetic field strength $B$,
such as $L \sim 30 \, (B/10^{-9}{\rm Gauss})^3 \rm pc$.  The resulting
CMB temperature and polarization anisotropies for the initial power
law index of the magnetic fields $n>3/2$ are somewhat different from
the ones previously obtained by using linear perturbation theory. In
particular, differences can appear on intermediate scales $l < 2000$
and small scales $l > 20000$.  On scales $l < 2000$ the CMB anisotropy
and polarization spectra are flat in the case of our nonlinear
calculations whereas the spectra have a blue index calculated with
linear perturbation theory if we assume the velocity fields of baryons
induced by the magnetic fields achieved Alfv\'en velocity due to the
turbulent motions on large scales in the early universe.  Our
calculation gives a constraint on the magnetic field strength in the
intermediate scale of CMB observations. Upper limits are set by WMAP
and BOOMERANG results for comoving magnetic field strength of $B < 28
\, \rm nGauss$ with a comoving coherence length of $L > 0.7 \,\rm Mpc$
for the most extreme case, or $B < 30 \, \rm nGauss$ and $L> 0.8 \,
\rm Mpc$ for the most conservative case.  We may also expect higher
signals on large scales of the polarization spectra compared to linear
calculations.  The signal may even exceed the B-mode polarization from
gravitational lensing depending on the strength of the primordial
magnetic fields.  On very small scales, the diffusion damping scale of
nonlinear calculations turns out to be much smaller than the one of
linear calculations if the comoving magnetic field strength $B > 16
\, \rm nGauss$. If the magnetic field strength is smaller, the
diffusion scales become smaller too.  Therefore we expect to have
both, temperature and polarization anisotropies, even beyond $l
>10000$ regardless of the strength of the magnetic fields.  The peak
values of the temperature anisotropy and the B-mode polarization
spectra are approximately $40 \, \mu \rm K$ and a few $\mu \rm K$,
respectively.

\end{abstract}

\maketitle

\section{Introduction}

Magnetic fields have been observed in many galaxies and galaxy
clusters \cite{obmag, maggalaxy}.  Observations have revealed that
these magnetic fields typically have a few $\mu $Gauss strengths and
relatively large coherent scales, i.e., a few tens of kpc for clusters
of galaxies and a few kpc for galaxies.  It is one of the great
challenges for modern astronomy to understand the origin of these
magnetic fields.

Perhaps the most conventional scenario of generating such magnetic
fields is as follows.  First, small seeds of the magnetic fields are
produced due to the Biermann battery mechanism.  Although the
resultant magnetic fields are very weak, those are amplified by the
dynamo process \cite{dynamo} (for a comprehensive review see
\cite{magreview}).  Eventually these magnetic fields are spread by
Supernova winds or AGN jets into inter-galactic medium.

However, the fact that magnetic fields with very large coherent scales
are observed in galaxy clusters or high redshift galaxies with a
strength of a few $\mu$Gauss casts some questions on this standard dynamo
scenario \cite{obmag, maghighz}: how can these coherent magnetic
fields be spread into inter-cluster medium, and is it possible for the
dynamo amplification to take place fast enough in high redshift
galaxies?  An alternative possibility to the dynamo scenario is the
generation of magnetic fields in the early universe. There are in fact
many previous works which suggest the generation of the magnetic
fields in the early universe, e.g. during an inflation or at cosmological phase
transitions (QCD or electroweak phase transition). For a detailed
review, see \cite{maggenereview}.

If magnetic fields are generated in the early universe, we need to
understand their evolution in the expanding hot universe to obtain
their field strength and structure at present. Several studies of
evolution of magnetic fields have been performed by employing the
linearized equations of magneto-hydro dynamics (MHD)~\cite{j-k-o,s-b-nonlinear}.
These studies found that the growth of magnetic fields in the early
universe is rather complicated such that the cosmological viscosity
plays an important role~\cite{j-k-o}. Magnetic field energy is
dissipated by the viscosity due to neutrinos and photons, equivalent
to Silk damping \cite{silk} for density fluctuations of baryon-photon
fluid. However, the damping efficiency is different for different MHD
modes. The Alfv\'en and the slow modes are damped less than the fast
modes so that the Alfv\'en and the slow modes survive at small scales
where the dissipation is effective. Meanwhile, it is also expected
that the nonlinear effects contribute to the evolution of magnetic
fields since equipartition between the magnetic field and the fluid
will be established and the magnetic fields will cascade from large scales
to small scales.  Nonlinear effects on Alfv\'en modes in the
presence of viscosity is investigated analytically for particular
configurations \cite{s-b-nonlinear} and only a little damping of 
Alfv\'en modes is found in these situations.

Recently, Banerjee and Jedamzik~\cite{b-j-letter,b-j-paper} studied
the evolution of Alfv\'en modes using MHD simulation in the expanding
universe including dissipation due to diffusion and
neutrino/photon-drag due to free streaming neutral particles. The
evolution of the primordial magnetic fields is solely determined by
the kinetic Reynolds number, which is defined by $R \equiv v^2 /L f$
where $v$ is the fluid velocity, $L$ is the length scale and $f$ is
the fluid dissipation (which is discussed in more detail in the
following section),
can be divided into three different regimes: turbulent regime $(R \gg
1)$, viscous (diffusion) regime and (viscous) free streaming regime $(R <
1)$. Furthermore, analytic expressions for the growth of the magnetic
coherence length due to small wavelength damping and helicity
conservation were found. Under optimistic assumptions it is possible
for magnetic fields produced during a QCD phase transition with an
initial coherence length of $1$ pc to have attained a kpc coherence
length at present.

The existence of the primordial magnetic fields leaves traces on
various cosmological phenomena.  Investigating the cosmological
effects of the magnetic fields in the early universe, i.e., the
effects on Big Bang nucleosynthesis (BBN), structure formation and
cosmological microwave background (CMB) anisotropies, we  can
constrain the strength and the structure of primordial magnetic
fields. 

Among them, the traces of primordial magnetic fields on CMB
anisotropies and polarization are of particular interest since they
provide information of not only the magnetic field energy but also the
coherence length.  On the contrary, BBN gives only a limit on the
magnetic field energy density ($B \lesssim 7 \times 10^{-5}$Gauss at
present)~\cite{bbnconstraint}.

The coherence length is an especially important observational quantity
if one wishes to understand the origin of cosmic magnetic fields
because the length scale strongly correlates with the production
mechanism.  The primordial magnetic fields produced by a causal
mechanism during, for example, a phase transition are limited by the
length scale which corresponds to the horizon scale at the epoch of
the magnetic filed generation.  In contrast, the magnetic fields
produced in the inflation epoch are expected to have the
scale-invariant spectrum such that the amplitudes of the magnetic
fields at the horizon crossing are the same over the all scales.

There have been already several works to set constraints on the
primordial magnetic fields by using current CMB data. But so far, non
of these considered the full non-linear evolution of Alfv\'en
modes. Barrow et al.~\cite{b-f-s} set a limit on the homogeneous
(coherent) magnetic field by using COBE data.  Such a homogeneous
magnetic field produces large scale anisotropic pressures and these
pressures require an anisotropic gravitational field to support them.
This anisotropic gravitational field should generate anisotropies in
CMB temperature.  They obtain a limit of $B \lesssim
10^{-9}$Gauss~\cite{b-f-s} at present.

Primordial magnetic fields with a coherence length comparable to the
horizon size at the last scattering surface (LSS) affect the sound
speed of the baryon-photon fluid and change their acoustic
oscillations.  This effect might be observed as the modification of
the acoustic peaks in the CMB angular power spectrum.  Adams et
al.~\cite{a-d-g} found that this modification should be detectable by
WMAP and PLANCK if the today's magnetic fields are $B>10^{-8}$Gauss
with the coherence length larger than the horizon at recombination.

A distortion of the CMB energy spectrum from the black body shape also
gives a constraint on primordial magnetic fields.  The magnetic
fields with a small coherent scale are dissipated due to viscosity
of the baryon-photon fluid and this dissipation energy distorts the
black body spectrum of CMB~\cite{diss-j-k-o}.  The spectrum
distortion is described by either a chemical potential $\mu$ or a
Compton $y$-parameter.  If the dissipation process takes place in the
very early universe when thermal equilibrium is maintained, it does
not affect the CMB spectrum.  As the temperature of the universe goes
down, however, thermal equilibrium can no longer be maintained.
Instead, kinetic equilibrium is preserved since only the Compton
scattering process which conserves photon number is effective.
If dissipation occurs during this stage, we expect to have $\mu$
distortion.  As the temperature drops more, eventually even kinetic
equilibrium can no longer be maintained and we expect to have $y$
distortion due to the dissipation.  COBE-FIRAS gives the CMB chemical
potential constraint of $\mu \lesssim 9 \times 10^{-5}$.  This limit
corresponds to the magnetic field constraint of $B \lesssim 3 \times
10^{-8}$Gauss at present on a comoving scale of $400$pc at redshift $z
\gtrsim 2 \times 10^6$~\cite{diss-j-k-o}.  On the other hand, the constraint by the
Compton $y$-parameter, which is $y \lesssim 1.5 \times 10^{-5}$, leads
to the constraint on the magnetic fields of $B \lesssim 3 \times
10^{-8} $ Gauss on 600kpc.

Another important constraint can be given from small scale CMB
temperature anisotropies and polarization.  Magnetic fields induce
peculiar velocities in photon-baryon fluids driven by Alfv\'en modes.
These peculiar velocities generate temperature anisotropies and
polarization in CMB. Observations by WMAP already constrain
these anisotropies induced by Alfv\'en modes to be smaller than the
primary anisotropies due to acoustic oscillations of the photon-baryon
fluid.  Therefore, one might conclude that there is no chance to
measure magnetic field sourced temperature fluctuations. However this
is not the case for small scale anisotropies.  Primary CMB
anisotropies and polarization at the recombination epoch are expected
to have a lack of small scale power due to the diffusion damping of
the photon-baryon fluid (Silk damping).  On the other hand,
calculations of linear evolution of Alfv\'en modes show less
damping~\cite{s-b-temp,m-k-k,al,y-i}.  These Alfv\'en modes generate
CMB temperature anisotropies and polarization.  Therefore we can
likely expect to have dominant contribution on small scale CMB
anisotropies from the magnetic fields if they exist.
 Using the WMAP result and taking into account both scalar 
  and Alfv\'en modes, one can constrain primordial magnetic fields to be
$B \lesssim 3.9 \times 10^{-9}$Gauss on 1Mpc at present by using the
Markov Chain Monte Carlo method~\cite{y-i}.
 
For the same reason, polarization induced by Alfv\'en modes from
primordial magnetic fields may dominate on small
scales~\cite{s-s-poral}. It is known that the polarization can be
decomposed into two modes, i.e., the E-mode polarization (gradient
component) and the B-mode polarization (rotation component) \cite{s-z}.
Observationally they are distinguishable. It is very interesting that
Alfv\'en modes only produce B-mode polarization since its
perturbations are of vector type.  On the contrary, scalar type
perturbations, which form the large scale structure of the universe,
only produce E-mode polarization.  It has been known that the
gravitational lensing effect on the primary E-mode polarization caused
by structure of the universe produces B-mode polarization on
intermediate scales (a few tens of arc minute angular size).
However, polarization of the magnetic field origin can be dominant as
a B-mode on small sales.


A few authors have investigated the generation of B-mode polarization
either by an analytic treatment with employing the tight coupling
approximation \cite{s-b-poral} or by direct numerical calculations
\cite{m-k-k,al}.  For the evolution of Alfv\'en mode perturbations,
however, they all apply linear analysis.  In this paper, we study the
evolution of Alfv\'en modes using recent MHD simulation by Banerjee
and Jedamzik~\cite{b-j-letter,b-j-paper}, which allows us to incorporate the
fully non-linear and self consistent spectra of the Alfv\'en modes to
calculate resulting CMB temperature anisotropy and polarization
spectra.

This paper is organized as follows.  In Sec. II, we summarize the
evolution of Alfv\'en modes based on the numerical MHD simulation of
Banerjee and Jedamzik.  We separate the evolution into three regimes,
i.e., turbulent, viscous and free streaming regimes.  In each regime,
the evolution of velocity fields is carefully investigated.  In
Sec. III, the power spectrum of the velocity fields is calculated.
Using this power spectrum, in Sec. IV, we compute CMB temperature
anisotropy and polarization spectra.  Sec. V is devoted to discussion,
in which we compare our results to those obtained with linear
perturbation theory. We give our conclusions in Sec. VI.  Throughout
the paper, we take WMAP values for the cosmological parameters, i.e.,
$h=0.71 \ (H_0=h \times 100 {\rm Km/s \cdot Mpc})$, $T_0 = 2.725$K,
$\Omega _{\rm b} h^2 =0.0224$ and $\Omega_{\rm M} h^2 =0.135$
\cite{wmap}.

\section{Evolution of Alfv\'en modes}
\label{nonlinear evolution}

Our final goal is to investigate the effect of the primordial magnetic
fields on the CMB temperature anisotropies and polarization.
Magnetic stresses produce vortical modes, such as the
``Alfv\'en mode'' and "slow magnetosonic mode" in the ionized fluids. 
As both are very similar we will refer to them as Alfv\'en modes, 
henceforth. These Alfv\'en modes
generate additional temperature anisotropies and polarization by
Doppler shift.
Therefore, we need to know the evolution of Alfv\'en
modes first.
Unlike previous works which employed a linear
perturbation approximation~\cite{s-b-temp,s-b-poral,m-k-k}, Banerjee
and Jedamzik~\cite{b-j-letter,b-j-paper} recently investigated the
nonlinear evolution of magnetic fields and Alfv\'en modes using
numerical MHD simulation. 
These simulations cover the three different
damping regimes appearing in the early universe: turbulence, viscous,
and free streaming. Based on their work, we
summarize the evolution of the velocity fluctuations (Alfv\'en modes)
of the ionized fluid in this section.

The growth of Alfv\'en modes is affected by the interaction
with photons and/or neutrinos in the early universe.  Here, 
we consider the interaction with photons only as neutrinos are already
decoupled from the cosmic evolution at the epoch of LSS.

The MHD equations in the expanding universe including diffusion due to
the photon background are given by~\cite{b-j-paper}
\begin{equation}
\frac{\partial {\bf v}}{\partial t}+ \frac{1}{a} ({\bf v} \cdot
\nabla){\bf v} +(1-3c_{\rm st} ^2)H {\bf v} +\frac{1}{a}
\left(\frac{{\bf B}\times(\nabla \times {\bf B})}{4 \pi
(\rho_{\rm t}+p_{\rm t})}\right) ={\bf f},
\label{eular} 
\end{equation}

\begin{equation}
\frac{\partial {\bf B}}{\partial t} +2H  {\bf B}
=\frac{1}{a}\nabla \times ({\bf v} \times {\bf B}),
\label{induction}
\end{equation}
where $\rho_{\rm b}$ is the baryon density, $\rho_{\rm t}$, $p_{\rm t}$ and $c_{\rm st}$
denote the
total density, pressure and sound velocity of photons and  baryons, $H$ is the
Hubble parameter and ${\bf f}$ is the dissipation term.  
The velocity ${\bf v}$ of photon-baryon fluids  here is the Alfv\'en
mode, 
and as such is incompressible so that ${\bf v}$ satisfies $\nabla \cdot {\bf v} =0$.  The
dissipation term, ${\bf f}$, is written as
\begin{equation}
{\bf f}=
a^{-2} \nu (\rho +p)^{-1} (\nabla^2 {\bf v} ),
\end{equation}
where the shear viscosity $\nu$ 
is given by \cite{w-viscous}
\begin{equation}
\nu=\frac{4}{15}\rho_{\gamma} L_{\rm mfp} .
\label{visco}
\end{equation}
Here $\rho_{\gamma}$ is the radiation energy density and $L_{\rm mfp}$
is the photon mean free path, $L_{\rm mfp} = (\sigma_{\rm T} n_e)^{-1}$
where $\sigma_{\rm T}$ is the Thomson cross section and $n_e$ is the free
electron density.

By using  $\nabla \cdot {\bf B} =0$,
Eq. (\ref{eular}) is rewritten as
\begin{equation}
\frac{\partial {\bf v}}{\partial t}+
\frac{1}{a} ({\bf v} \cdot \nabla){\bf v}+(1-3c_{\rm st} ^2) H {\bf v} 
-\frac{1}{a} ({\bf v}_{\rm A} \cdot \nabla) {\bf v}_{\rm A}  ={\bf f},
\label{eular2}
\end{equation}
where ${\bf v}_{\rm A}$ is the Alfv\'en velocity, ${\bf v}_{\rm A} \equiv {\bf B}/
\sqrt{4 \pi (\rho_{\rm t} + p_{\rm t})}$.

The dissipation term, ${\bf f}$ which assumes multiple scattering between
photons and baryons is valid only on scales larger than $L_{\rm
mfp}$.  Since $L_{\rm mfp}$ evolves rapidly as $a^3$, it soon exceeds
the wave length of a particular mode we consider. In other word, 
the comoving wave number of the velocity, $k$, becomes 
$a/k < L_{\rm mfp}$.  
Subsequently, we need to treat baryons (ionized fluids) and
photons separately.

Once the diffusive description becomes invalid, Alfv\'en modes will be
damped due to free streaming background photons. Note that the damping
time on the integral scale in the viscous regime is longer
than the Hubble time and dynamic evolution of the magnetic field is
'stalled'.  In the free streaming regime, the dissipation process can
be described as radiation drag whose force is proportional to the
fluid velocity ${\bf v}$.
The coefficient $\alpha$ of the drag force is given by \cite{p-fs}
\begin{equation}
\alpha = \frac{4}{3} \frac{\rho_{\gamma}}{\rho_{\rm b}}L_{\rm mfp} ^{-1} .
\label{alpha_drag}
\end{equation}
Therefore, the Euler equation for the baryon fluid with free
streaming photons is given by
\begin{equation}
\frac{\partial {\bf v}}{\partial t}+ \frac{1}{a} ({\bf v} \cdot
\nabla){\bf v}+ H {\bf v} -\frac{1}{a}\left( {\rho_{\rm t} +p_{\rm t} \over
\rho_{\rm b}}\right) ({\bf v}_{\rm A} \cdot \nabla) {\bf v}_{\rm A} =- \alpha {\bf v}.
\label{baryoneular}
\end{equation}

Let us now consider a characteristic scale on which most of the
magnetic energy exists.   This characteristic scale 
corresponds to the peak of the magnetic field power spectrum 
and can be defined by using the two point correlation function 
$
\xi (r) = \left\langle |{\bf B (x+r) B(x)} |\right\rangle 
$
as
\begin{equation}
L_{\rm int} = {1 \over \xi (0)}{ \int^{\infty}_{0} dr \xi (r)} .
\end{equation}
We refer to this scale as the {\em integral scale} and define the
comoving wave number of the integral scale as 
$k_{\rm int} \equiv a/L_{\rm int}$.

The integral scale grows in time. In the turbulent regime a direct
cascade due to non-linear interactions transfers energy from the
integral scale to the much smaller viscous scale, where it is lost to
heat.
In the free streaming regime the integral scale may also grow
due to the dissipation of flows on the integral scale itself.
In both regimes the damping time scale
is given by
\begin{equation}
t_{\rm eddy } \approx {L \over v} \approx {a \over kv} .
\label{eddytimescale}
\end{equation}
By comparing this eddy time scale on the integral scale with the
cosmic time $H^{-1}$, we can judge whether non-linear cascade
processes are important or not.  
If once $t_{\rm eddy }^{\rm int} \equiv L_{\rm int}/v < H^{-1}$ 
is satisfied, magnetic energy dissipation becomes effective and 
magnetic energy on scales $L\, <\, L_{\rm int}$ is lost.  

Now let us classify the evolution of the MHD nonlinear Alfv\'en modes into
three regimes, i.e., turbulent, viscous, and free streaming regimes,  
following the results of the numerical simulation.  
In the following argument, we mostly consider the evolution of 
Alfv\'en modes at the integral scale.

\subsection{Turbulent regime}\label{velocity_turb}

In the early universe, since 
the photon mean free path $L_{\rm mfp}$, which is proportional to 
$1/ n_e$, is very short, the effect of 
viscosity on Alfv\'en modes at the integral scale
is negligible until the mean free path becomes large enough 
to have Reynolds number $R\sim 1$ on $L_{\rm int}$. For $R>1$ we can
ignore the dissipation term 
${\bf f}$ in Eq.~(\ref{eular2}).  The advective term 
$({\bf v}_{\rm A} \cdot \nabla) {\bf v}_{\rm A} / a$ drives the fluid velocity in
this regime, which we call the turbulent regime.  

Ignoring the cosmological expansion term,  we 
find that the fluid velocity eventually approaches an equipartition state:
\begin{equation}
{\bf v}= {\bf v}_{\rm A}.
\label{tur-velo}
\end{equation}
This behavior is consistent with the MHD numerical simulation
\cite{b-j-letter,b-j-paper}.  
 Note that it is not clear whether the fluid velocity approaches an 
equipartition state on the  scales larger than the integral scale or
not since it takes longer on larger scales.  We will discuss this
point in Section 3.~A.

 If magnetic fields do not
decay, the Alfv\'en velocity stays constant in time since 
$B \propto a^{-2}$ for the adiabatic expansion and $\rho_{\rm t} \propto a^{-4}$ during
the radiation dominated epoch. Accordingly the eddy time scale at the
integral scale evolves as 
$
t_{\rm eddy }^{\rm int}  
=L_{\rm int}/v_{\rm A} \propto L_{\rm int}/\left( B/\sqrt{\rho_{\rm t}} 
\right)   \propto a . 
$ 
On the other hand, the cosmic time $1/H$ is proportional to $a^{2}$ in
the radiation dominated epoch.  Therefore soon or later the cosmic
time exceeds the eddy time, which allows for turbulent decay of the
magnetic fields.

Let us go into details about the evolution of the integral scale.  By
definition the integral scale corresponds to the peak location of the
energy power spectrum of the magnetic field.  Correspondingly, the
Alfv\'en velocity, which is proportional to the amplitude of the
magnetic field, peaks at the integral scale. Since we assume a blue
magnetic power spectrum, i.e $n > 0$, the eddy turnover time on the
scales larger than the integral scale is longer than $t_{\rm eddy
}^{\rm int}$.  Therefore the condition $t_{\rm eddy} = 1/H$ is first
satisfied on the integral scale and then gradually moves towards
larger scales.  Meantime, the cascading decay of the short wavelength
modes ($k > k_{\rm int}$) shifts the integral scale towards larger
scales.

The underlying physical process of the cascade decay is following.
When nonlinear effects become prominent, the magnetic field energy and
the fluid kinetic energy achieve equipartition.  The flow
eddies break into the smaller eddies.  Hence the kinetic energy is
transported from large scales to small scales by a nonlinear cascade
(Kolmogorov process).  The transported energy is changed into heat at
the scale where the dissipation process is effective.  Consequently,
the magnetic field energy at the integral scale is converted into heat
which is called direct cascade.




\subsection{Viscous regime }\label{velocity_visc}

As the universe evolves, the photon mean free path, $L_{\rm mfp}
\propto a^3$, becomes larger.  Dissipation due to photon interaction
becomes efficient and velocity fluctuations are damped by the photon
drag.  Eventually the dissipation term dominates the advective term
at the integral scale in Eq. (\ref{eular2}).  This
is the second regime, which is refereed as the viscous regime.  During
the viscous regime, the eddy time at the integral scale is always
larger than the cosmic time due to the decay of the fluid velocity
$v$, which makes the eddy time longer. 

The transition epoch from turbulent to viscous regimes on the integral
scale can be determined by comparing the advective term 
$({\bf v} \cdot \nabla) {\bf v} / a$ and the dissipation term 
${\bf f} = \left(\nu /(\rho_{\rm t} + p_{\rm t})\right) \nabla^2 {\bf v} /a^2$.  
During the
turbulent regime, the amplitude of the fluid velocity in the advective
term is equal to the Alfv\'en velocity $v_{\rm A}$ as shown in the previous
subsection.  Therefore the advective term can be written by using the
integral scale as $({\bf v} \cdot \nabla) {\bf v} / a \simeq
v_{\rm A}^2/L_{\rm int}$.  The dissipation term can be also rewritten as $
\left( \nu /(\rho_{\rm t} + p_{\rm t})\right) v_{\rm A} /L_{\rm int}^2. $ At the
transition epoch, these two terms become equal, which yields 
\begin{equation}
v_{\rm A}={\nu \over (\rho_{\rm t} + p_{\rm t})L_{\rm int}} = {L_{\rm mfp}\over 5 L_{\rm int}} . 
\label{diffusion_nonlinear}
\end{equation}
Here we assume the radiation domination, i.e., $\rho_{\rm t} + p_{\rm t} =
4\rho_\gamma / 3$.
Until the beginning of the viscous regime, 
the eddy time $t_{\rm eddy} = L_{\rm int}/v_{\rm A} $ is equal to the
cosmic time $1/H$, which leads to 
$v_{\rm A} \simeq \sqrt{H/(n_e \sigma_{\rm T})}$. 
Inserting the definition of $v_{\rm A}$ into
this equation, the transition redshift $z_{\rm t-v}$ can be obtained as
\begin{equation}
z_{\rm t-v} \simeq 6\times 10^{6} B_{-9}^{-2} 
\simeq  6\times 10^{6} \left({k_{\rm int} \over 3.4 \times 10^4 {\rm Mpc}^{-1} }\right)^{2/3},
\label{redshift_tur_vis}
\end{equation} 
where $B_{-9}$ is the comoving magnetic field normalized by 
$10^{-9}$Gauss, i.e.,  $B_{-9}=(B/10^{-9})a^2$ and 
\begin{equation} 
k_{\rm int} \simeq  3.4 \times 10^4  B_{-9}^{-3}{\rm  Mpc^{-1}} ,
\label{eq:int_viscus}  
\end{equation} 
is the comoving wave number of the integral scale at the transition epoch.  
Note that Eqs.~(\ref{redshift_tur_vis}) and (\ref{eq:int_viscus}) are
only valid after $e^{\pm}$ annihilation where $L_{\rm mfp} =
1/\sigma_{\rm T}\,n_e$, i.e., $z < 10^8$. For earlier epochs
Eq.~(\ref{diffusion_nonlinear}) must be evaluated numerically to find
the redshift--B-field relations.


In the viscous regime, we can ignore the advective term in 
Eq.~(\ref{eular2}) at the integral scale since the fluid velocity 
$v$ decays due to the dissipation.    We can also omit the expansion 
term since the evolution of the fluid velocity is controlled by the 
dissipation whose time scale is much faster than the cosmic expansion.
Employing the terminal-velocity approximation, we obtain 
\begin{equation}
{v_{\rm A}^2 \over L_{\rm int} }= {\nu \over (\rho_{\rm t}+p_{\rm t})} 
{v \over L_{\rm int}^2}.  
\end{equation}
Using the comoving wave number at the integral scale $k_{\rm int}$, 
the fluid velocity can be described as 
\begin{equation}
v=(\rho _{\rm t}+ p_{\rm t}) \frac{v_{\rm A}^2}{\nu} \frac{a}{k_{\rm int}} ~ .
\label{viscous-velo} 
\end{equation}
Hence it is found that the evolution of the fluid velocity becomes 
$v \propto a^{-2}$.


According to this solution, the eddy time at the integral scale is
proportional to $a^3$ in both radiation and matter dominated epochs.
Therefore the eddy time remains longer than the cosmic expansion time,
$1/H$ which is proportional to $a^2$, and $a^{3/2}$ in the radiation
dominated epoch and the matter dominated epoch, respectively.  As a
result, no direct cascade occurs during the viscous regime and the
integral scale does not grow.

\subsection{Free streaming regime}\label{velocity_free}

 The third regime is the free streaming regime, in which photon mean
free path is larger than the integral scale.  In this regime, photons
and baryon fluids are decoupled and magnetic fields can amplify the
fluid velocity.  This amplification of the velocity makes the eddy
time shorter until the eddy time becomes equal to the cosmic time
$1/H$.  At this point the kinetic energy on $L_{\rm int}$ is directly
dissipated into heat and the integral scale shifts to larger scales.

Let us first estimate the transition epoch from the viscous regime to
the free streaming regime.  The transition happens when the mean free
path $L_{\rm mfp}$ becomes equal to the integral scale $L_{\rm int}$.
Employing the comoving wave number of the integral scale in the viscous
regime Eq.~(\ref{eq:int_viscus}), we obtain the transition redshift as
\begin{equation}
z_{\rm v-f} \simeq 1.7\times 10^{5} B_{-9}^{-3/2} \simeq 
1.7 \times 10^5 \left( k \over 3.4\times10^4 {\rm Mpc}^{-1} \right)^{1/2}, 
\label{redshift_vis_free}
\end{equation}

Using Eq.~(\ref{viscous-velo}), and noting that $a/k_{\rm int}\approx
L_{\rm mfp}$ during this epoch, we find that $v$ has decayed to
$v\approx v_A^2$ at the transition.  In the free streaming regime, we
need to solve Eq.~(\ref{baryoneular}) instead of Eq.~(\ref{eular2})
for the evolution of the fluid velocity.  Ignoring the cosmological
expansion term, we obtain the solution with the terminal velocity
approximation as
\begin{equation}
v={({\rho_{\rm t} +p_{\rm t}}) k_{\rm int} v_{\rm A} ^2 \over {\rho_{\rm b}} \alpha a}.
\label{fs-velo}
\end{equation}

The eddy time is longer than the cosmic time when the free streaming
epoch begins since $v$ is much smaller than $v_{\rm A}$ during the viscous
regime due to dissipation.  From the above solution, however, it is
found that $v \propto a^2$ in the free streaming epoch.  Accordingly
the eddy time evolves as $a^{-1}$.  Therefore the eddy time soon
becomes shorter than the cosmic time which is proportional to $a^{2/3}$
in the matter dominated universe.  

As is shown in the previous subsection, the integral scale does not
evolve in the viscous regime. In the free streaming regime, the
integral scale does also not change until the eddy time becomes equal
to the cosmic time.

When the eddy time is equal to the 
cosmic time on the integral scale, i.e. $L_{\rm int}/v = 1/H$,
the comoving wave number must satisfy the relation (using
Eqs.~(\ref{fs-velo}) and (\ref{alpha_drag})):
\begin{equation}
k_{\rm int} ^{-1} \simeq  {v_{\rm A} \over k_{\rm S}}, 
\label{eq:int_free}
\end{equation}
where $k_{\rm S}^{-1}$ is the comoving Silk scale defined as $k_{\rm
S}^{-1} \simeq \sqrt{L_{\rm mfp}/H}/a$.  Note that in this regime,
$v_{\rm A}/k_{\rm S}$ is the comoving damping scale for Alfv\'en
modes.  Unlike in the non-magnetized photon-baryon fluid overdamped
Alfv\'en modes survive below the Silk damping scale for weak magnetic
fields as was first pointed out by~\cite{j-k-o} and~\cite{s-b-nonlinear}.  While
it is not clear either the direct cascade or the diffusion process
takes place first, the baryon velocity exponentially damps away bellow
this scale.

Substituting Eq.~(\ref{eq:int_viscus}), which is the 
comoving integral wave number in the viscous regime and in the free
streaming regime until the damping scale for Alfv\'en modes, $v_{\rm A} / k_{\rm
  S}$, becomes equal to the integral scale, into 
Eq.~(\ref{eq:int_free}), 
we obtain the redshift below which further growth of the integral scale 
in the (viscous) free streaming regime occurs 
\begin{equation}
z = 1100 \left( {B_{-9} \over 16 {\rm nGauss}} \right)^{-8/5}
=1100\left({k_{\rm int} \over 8 {\rm Mpc}^{-1}}\right)^{8/15}.  
\label{eq:free_decay}
\end{equation}
Therefore if the integral scale once became larger than $100$kpc by the 
end of the turbulent regime,  the integral scale 
did not change through viscous and free streaming regimes all the way 
until the recombination epoch.  

On the contrary, if the integral scale was smaller than 
$100$kpc,  there was further growth of $L_{\rm int}$ during the free
streaming regime before recombination.  The integral scale shifts to a larger 
scale in this case.  The resultant integral scale at the recombination 
epoch depends on the slope of the magnetic field power spectrum.

In Fig.~1 we show the areas of turbulent, viscous and free streaming regimes
in the $z-k_{\rm int}$ plane.  
Here we employ Eqs. (\ref{redshift_tur_vis}), (\ref{redshift_vis_free}),
and (\ref{eq:free_decay}).
Knowing the time evolution of the integral scale $k_{\rm
int}$, which depends on the magnetic field spectrum, we can draw an
evolutionary track in this plane and easily understand when the
transition between different regimes occurs.  We will discuss this
evolution in the next section.

\begin{figure}[htbp]
  \begin{center}
    \includegraphics[keepaspectratio=true,height=70mm]{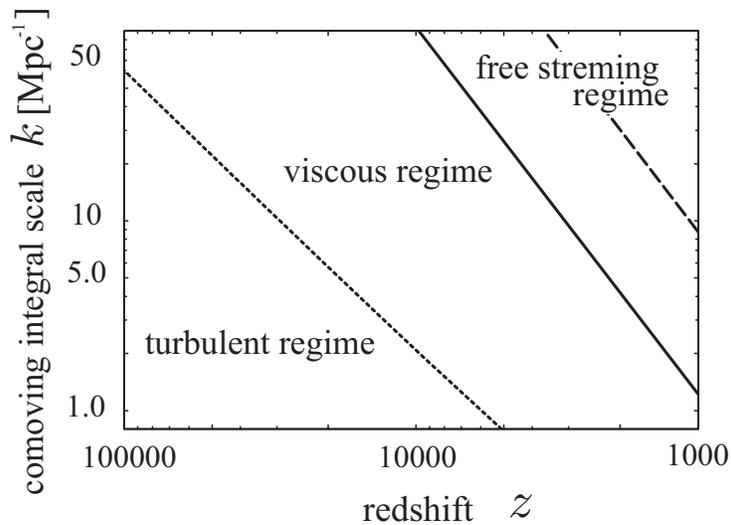}
  \end{center}
  \caption{The transition comoving scales as the function of the redshift.
  The dotted line is the transition line from the turbulent to the viscous regime,
  Eq.~(\ref{redshift_tur_vis}).
  The solid line is that from the viscous to the free streaming regime,
  Eq.~(\ref{redshift_vis_free}). 
  The dashed line represents the start of the direct cascade 
  in the free streaming regime, Eq. (\ref{eq:free_decay}), which is 
  more or less when the diffusion process takes place.
  In the left side of this line there is no direct cascade or
  diffusion damping.
  In the right side the direct cascade or diffusion damping is occurring.}
  \label{fig:z-vis-fs}
\end{figure}

\section{Alfv\'en Mode Spectrum}\label{power-spectrum-alfven}

In the previous section, we discussed the evolution history of magnetic
fields and fluid velocity or Alfv\'en modes.  We separated the
evolution into three regimes, i.e., turbulent, viscous, and free
streaming.  We analytically solved the MHD-Euler equation and acquired
the relation between the fluid velocity $v$ and the Alfv\'en velocity
$v_{\rm A}$ in each regime.  Once $v$ was obtained, we could estimate
the eddy time $t_{\rm eddy}$ and compare it with the cosmic time 
in order to know whether field evolution occurred.

However we can not describe the time evolution of the integral scale
of the magnetic fields without knowing the magnetic field power
spectrum.

In this section, we find the power spectra of
magnetic fields and the fluid velocities for a given
initial shape of the magnetic field power spectrum.
Being motivated by a causal mechanism of primordial magnetic field
generation such as QCD phase transition, we employ the power law
spectrum with exponential cutoff for initial comoving magnetic fields
as
\begin{equation}
{\cal P}_{B}^{\rm comov} (t_{\rm i}, k)={k^3 \over 2 \pi^2} 
\langle |B_{\rm comov}(t_{\rm i} , k)|^2
\rangle = B_{\rm comov}^2 \left(k \over k_{\rm c} \right)^n 
\exp \left(-\left(k \over
k_{\rm c} \right)^2 \right) , 
\label{initial-mag-powerlow}
\end{equation}
where $n$ is the initial power law index and $k_{\rm c}$ is the cutoff
scale, and $B_{\rm comov}$ is the amplitude of the initial comoving
magnetic fields at the cutoff scale.  Obviously, the cutoff scale
corresponds to the integral scale at the initial epoch for a blue, $n
> 0$ spectra.

To see the evolution of the fluid velocity induced by the magnetic
field, it is convenient to define the power spectrum of the Alfv\'en
velocity ${\cal P}_{\rm A}(k)$ as
\begin{equation}
{\cal P}_{\rm A}(t, k)
\equiv {k^3 \over 2\pi^2}  \langle \vert v_{\rm A} (t, k) \vert^2 \rangle
= {1 \over 4 \pi}{1 \over \left(\rho_{\rm t} + p_{\rm t}\right)a^4}
{\cal P}_B^{\rm comov}(t, k) .  
\end{equation}

\subsection{Turbulent regime}

In the turbulent regime, we have shown that the magnetic field energy
reaches an equipartition state with the fluid kinetic energy, i.e.,
$v=v_{\rm A}$ (Eq.~(\ref{tur-velo})).

Therefore we expect to have the following relation between the power
spectrum of the fluid velocity ${\cal P}_v(t,k)$ and the energy density
of the magnetic field, i.e. the power spectrum of the Alfv\'en velocity,
${\cal P}_{\rm A}(t,k)$, as
\begin{equation}
{\cal P}_{\rm A}(t,k)= P_v(t,k).  
\label{eq:alf&mag}
\end{equation}
Here 
the power spectrum of the fluid velocity is defined as 
\begin{equation}
{\cal P}_v(k)= {k^3 \over 2\pi^2} \langle |v|^2  \rangle .
\end{equation}

\begin{figure}[tbp]
  \begin{center}
 \includegraphics[keepaspectratio=true,height=65mm]{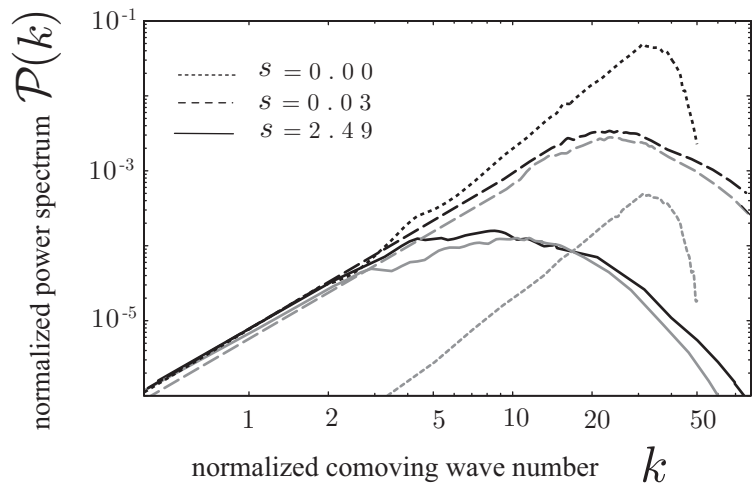}
 \caption{ The evolution of the comoving energy power spectrum of the
 magnetic fields and the kinetic energy power spectrum of the fluids
 in the turbulent regime from the numerical simulation by
 \cite{b-j-letter, b-j-paper}.  The x-axis is a comoving wave-number.
 The normalizations of x and y-axises are arbitrary.  The thin and
 thick lines are the power spectra of magnetic fields and fluids,
 respectively.  The dotted lines are the initial power spectra
 ($\tau=0.00$) with the spectrum index $n=4$.  This figure shows that
 the energy densities become the equipartition, $v=v_{\rm A}$ after a
 few eddy time scale even on large scales.}
  \label{fig:tur-power}
   \end{center}
\end{figure}

Eq.~(\ref{eq:alf&mag}) is consistent with the results of the MHD
simulation as is shown in FIG. \ref{fig:tur-power}.  In this figure
$s$ is the time stamp and has the following relation with the redshift
$z$,
\begin{equation}
\frac{z+1}{z_{\rm i} +1} =\exp \left(-s \frac{t_{\rm i \rm eddy}}{H_0
^{-1}}\right),
\label{power-t-def}
\end{equation}
where the subscript $\rm i$ denotes the initial value.  Initially, the power
spectrum of the fluid velocity is very different from the one of the
magnetic field.  However, the fluid velocity power spectrum soon
catches up with the magnetic field one.  
Subsequently, they evolve
in a very similar manner. In this figure, it is shown fluid velocity
on large scales also immediately reaches the equipartition as well as
the velocity on scales smaller than the integral scale.   
Therefore hereafter we assume the equipartition on all scales, i.e., 
Eq.~(\ref{eq:alf&mag}).  Although it is immediate in terms of $s$,
however, it may take longer time in physical time.  If this is the
case and the fluid velocity on larger scales has not caught up with the Alfv\'en
velocity by the end of the turbulent regime, 
we expect that the power spectrum of the fluid velocity 
${\cal P}_v(k)$ has less power on large scales than the one of the
magnetic fields ${\cal P}_{\rm A}(k)$ which is proportional to $k^n$ on
large scales.  
The most extreme case is that turbulence does not work at all on large
scales, in which we need to employ linear perturbation.
Subramanian and Barrow \cite{s-b-temp} obtain 
${\cal P}_v(k) \propto k^5$ for the linear calculation. 
Therefore the power law slope on large scales can be between $n$ and
$5$.

On scales larger than the integral scale, the power spectrum of the
Alfv\'en velocity keeps its initial amplitude.  
The power spectrum of the fluid velocity also soon reaches to this 
amplitude.

Let us now estimate the evolution of the integral scale.  
In Sec. \ref{velocity_turb},  
we explained the reason why this scale shifts to larger 
scale with time during the turbulent regime.  
Now we would like to estimate the rate of this shift as a function 
of the power law index $n$.  

As we have shown in \ref{velocity_turb}, the integral scale is the
scale where $t_{\rm eddy} = 1/H$ is satisfied.
In the turbulent regime
$t_{\rm eddy} = a/(k_{\rm int}v)=a/(k_{\rm int}v_{\rm A})$.
Therefore $k_{\rm int} = aH/v_{\rm A}$.  The time evolution of the
Alfv\'en velocity at the integral scale,
in the absence of dissipation, 
is estimated as 
$v_{\rm A}=B/\sqrt{\rho_{\rm t}+p_{\rm t}} 
\propto \sqrt{{\cal P}_{\rm A}(k_{\rm int})} \times a^0 $.
On the scale larger than the previous integral
scale, ${\cal P}_B(k) \propto k^n$.  
Therefore the growth of
the integral scale, 
when dissipation is also considered, 
is obtained as
\begin{equation}
k_{\rm int} \propto a^{-2/(n+2)} . 
\label{rate-tur}
\end{equation}
This analytic estimate is supported by the numerical simulation as is 
seen  in FIG. \ref{fig:tur-power}.  
This figure also shows the direct cascade of the magnetic fields 
soon smears out the initial sharp damping in the power spectrum.

The resultant power spectrum of the magnetic fields during the 
turbulent regime has 
power law shape on both large and small scales but with different 
indices.  Matching the amplitude with the initial value on  
large scales, we obtain 
\begin{eqnarray}
{\cal P}_{\rm A} (t,k) & = & {\cal P}_{\rm A} (t_{\rm i},k) 
\propto  k^n, \qquad  k<k_{\rm int} , \\ 
{\cal P}_{\rm A} (t,k)& = & {\cal P}_{\rm A} (t_{\rm i},k_{\rm int}) 
\left({k \over k_{\rm int}}\right)^m 
\propto  k^m, \  k>k_{\rm int}, 
\end{eqnarray} 
where $m$ as observed in the numerical simulation  
is approximately $m \sim -2/3 $ (for details see \cite{b-j-paper}). 

Because of the equipartition, the evolution of the power spectrum of
the fluid velocity is identical to the Alfv\'en velocity
as
\begin{eqnarray}
{\cal P}_v (t,k)  &=&  {\cal P}_{\rm A} (t_{\rm i},k) 
\propto  k^n, \qquad k<k_{\rm int} , \\ 
{\cal P}_v (t,k) &=&  {\cal P}_{\rm A} (t_{\rm i},k_{\rm int}) 
\left({k \over k_{\rm int}}\right)^m 
\propto  k^m, ~
k>k_{\rm int}  . 
\end{eqnarray}

\subsection{Viscous regime}\label{velo-viscoussection}

When the relation $v_{\rm A}=(\rho_{\rm t}+p_{\rm t})\nu k/a$ is satisfied at the integral scale,  
the magnetic fields enter into the viscous regime.
In the viscous regime 
the magnetic fields do not decay.
Therefore the spectrum of the magnetic fields is frozen  
over the viscous regime and keeps its feature at the transition 
epoch  from the turbulent to the viscous regimes $t_{\rm t-v}$ as 
\begin{equation}
{\cal P}_{\rm A} (t, k)= {\cal P}_{\rm A} (t_{\rm t-v}, k)  .
\end{equation}

In the viscous regime no further growth of $L_{\rm int}$ happens
and the amplitude of the fluid velocity is proportional to
square of the Alfv\'en velocity amplitude as is shown in
Eq. (\ref{viscous-velo}).  Note that Eq. (\ref{viscous-velo}) can be
applied not only to the integral scale but also to all scales during
the viscous regime.  Therefore the power spectrum of the fluid
velocity can be described using the power spectrum of
the Alfv\'en velocity.

If the power law index $n$ is smaller than $3/2$, the contribution from 
larger $k$ (smaller scales) is negligible and we obtain 
\begin{equation}
{\cal P}_v(t,k)
\approx \left(  {\left(\rho_{\rm t}+p_{\rm t}\right) a \over \nu k} 
{\cal P}_{\rm A}(t_{\rm t-v} ,k)\right)^2 
\propto 
\left \{
\begin{array}{ll}
k^{2(n-1)}, & \qquad k<k_{\rm int},\\
k^{2(m-1)}, & \qquad k>k_{\rm int}.
\end{array}
\right.
\label{power-velo-vis-red}
\end{equation}

If $n$ is larger than $3/2$, on the other hand, 
the contribution from the integral scale dominates in the convolution
on scales larger than the integral scale.  
Then we can simply describe the power spectrum of the fluid velocity as 
(cf. Appendix A, Eq.~(\ref{power-line-vis}) and text below)
\begin{eqnarray}
{\cal P}_v(t,k)
& \approx &
\left( {\left(\rho_{\rm t}+p_{\rm t}\right) a \over \nu k_{\rm int}} 
{\cal P}_{\rm A}(t_{\rm t-v} ,k_{\rm int}) 
\right)^2
\left({k \over k_{\rm int}}\right) \propto k
  , \quad k<k_{\rm int} ,
  \label{power-velo-vis-blue-large}\\
{\cal P}_v(t,k)
& \approx & 
\left( {\left(\rho_{\rm t}+p_{\rm t}\right) a \over \nu k} 
{\cal P}_{\rm A}(t_{\rm t-v} ,k)\right)^2 \propto k^{2(m-1)}
 , \quad k > k_{\rm int} . 
\label{power-velo-vis-blue}
\end{eqnarray}
Note that in Eq.~(\ref{power-velo-vis-blue-large}), the slope may
be steeper on large scales if the equipartition was not  achieved
during the turbulent regime.

On scales smaller than the photon mean free path $L_{\rm mfp}$, 
the diffusion approximation is no longer valid
as is the case in the free streaming regime regardless of the value of
the initial power law index $n$.
Therefore we shall use the argument of Sec. \ref{velocity_free}
for the evolution of the fluid velocity.  
Employing Eq. (\ref{fs-velo}), we obtain 
\begin{equation}
{\cal P}_v(t,k)
\approx 
\left( {\left(\rho_{\rm t}+p_{\rm t}\right) k \over \rho_{\rm b} \alpha  a} 
{\cal P}_{\rm A}(t_{\rm t-v} ,k)\right)^2 \propto k^{2(m+1)}
, \quad k > k_{\rm f}  ,  
\label{power-velo-vis-blue-small}
\end{equation}
where $k_{\rm f} \equiv \sqrt{5}a/L_{\rm mfp}$.  
For the definition of $k_{\rm f}$, we put the factor $\sqrt{5}$ 
in order to match the power spectrum at $k_{\rm f}$.  
The estimated value of $k_{\rm f}$ at the recombination epoch is
$k_{\rm f}=1.2 \rm Mpc ^{-1}(z/1100)^2$.

\subsection{Free streaming regime}\label{velo-fssection}

When the integral scale becomes smaller than the photon mean free
path, i.e., $k_{\rm f} = k_{\rm int}$, the free streaming regime
begins.  At the beginning of the free streaming regime, the eddy time
at the integral scale is longer than the cosmic time.  
Therefore the integral scale does not change.  If
the integral scale is larger than $100$kpc, the eddy time is always
longer than the cosmic time until recombination as is described in
Eq. (\ref{eq:free_decay}), 
and further dissipation never occurs.  In
this case, the Alfv\'en velocity $v_{\rm A}$ does not evolve and the
fluid velocity is expressed in terms of $v_{\rm A}$ as $v= (\rho_{\rm
t}+p_{\rm t})kv_{\rm A}^2/(\rho_{\rm b} \alpha a)$
(Eq. (\ref{fs-velo})).  Accordingly the power spectrum of the fluid
velocity can be described by the convolution of the power spectrum of
the Alfv\'en velocity.

Following the argument we made in the previous subsection, 
we obtain the power spectrum of the fluid velocity as 
\begin{equation}
{\cal P}_v(t,k)
\approx \left(  {\left(\rho_{\rm t}+p_{\rm t}\right) k \over \rho_{\rm b} \alpha a } 
{\cal P}_{\rm A}(t_{\rm v-f} ,k)\right)^2 
\propto 
\left \{
\begin{array}{ll}
k^{2(n+1)}, & \qquad k_{\rm f}<k<k_{\rm int},\\
k^{2(m+1)}, & \qquad k>k_{\rm int} ,
\end{array}
\right .
\label{power-velo-free-red}
\end{equation}
for the power law index $n$ being smaller than $3/2$. 

If $n>3/2$, on the contrary, we obtain 
\begin{eqnarray}
{\cal P}_v(t,k)
& \approx &
\left( {\left(\rho_{\rm t}+p_{\rm t}\right) k_{\rm int} \over \rho_{\rm b} \alpha a } 
{\cal P}_{\rm A}(t_{\rm t-v} ,k_{\rm int}) 
\right)^2
\left({k \over k_{\rm int}}\right)^{5}  
\propto k^5, \quad k_{\rm f}<k<k_{\rm int} ,
\label{power-velo-free-blue}\\
{\cal P}_v(t,k)
&\approx &
\left( {\left(\rho_{\rm t}+p_{\rm t}\right) k \over \rho_{\rm b} \alpha a } 
{\cal P}_{\rm A}(t_{\rm v-f} ,k)\right)^2 
\propto k^{2(m+1)}, \quad  k>k_{\rm int}. 
\label{power-velo-free-blue-small}
\end{eqnarray}

On the scales larger than the free streaming scale, i.e., $k<k_{\rm f}$, 
the evolution of the fluid velocity still follows the solution 
of the viscous regime.  Therefore 
\begin{eqnarray}
{\cal P}_v(t,k)
& \approx &
\left( {\left(\rho_{\rm t}+p_{\rm t}\right) a \over \nu k_{\rm int}} 
{\cal P}_{\rm A}(t_{\rm t-v} ,k_{\rm int}) 
\right)^2
\left({k \over k_{\rm int}}\right)  
\propto k, \quad k<k_{\rm f}, \quad n>3/2, 
\label{power-velo-free-blue-large} \\
{\cal P}_v(t,k)
& \approx & 
\left(  {\left(\rho_{\rm t}+p_{\rm t}\right) a \over \nu k} 
{\cal P}_{\rm A}(t_{\rm t-v} ,k)\right)^2 
\propto k^{2(n-1)}, \quad k<k_{\rm f} , \quad n<3/2 . 
\label{power-velo-free-red-large}
\end{eqnarray}
Note that in Eqs.~(\ref{power-velo-free-blue-large}) 
and (\ref{power-velo-free-red-large}), the slopes may
be steeper on large scales as is pointed out in the previous subsection.

Finally, in the case of the integral scale at the transition epoch
from the viscous to free streaming regimes being smaller than $100$kpc, 
the further processing of $L_{\rm int}$ begins at the redshift of
Eq.~(\ref{eq:free_decay}).  Once further decay takes place, the
integral scale shifts to larger scales.
Let us now
estimate the evolution of the integral scale as a function of the
initial power law index $n$.  The integral scale can be written as
$k_{\rm int} = aH/v$.  The time evolution of the fluid velocity at the
integral scale is estimated as $v= (\rho_{\rm t}+p_{\rm t})kv_{\rm
A}^2/(\rho_{\rm b} \alpha a) \propto k_{\rm int} {\cal P}_{\rm
A}(k_{\rm int}) a^{2}$.  On the scale larger than the previous
integral scale, ${\cal P}_{\rm A}(k) \propto k^n$.  Therefore the time
evolution of the integral scale is given by
\begin{eqnarray}
k_{\rm int} &\propto a^{-3/(n+2)} ~~~~ 
{\rm in\ the\ radiation\ dominant}, \nonumber \\ 
k_{\rm int} &\propto a^{-5/(2(n+2))}~~~ {\rm in\ the\ matter\ dominant} . 
\label{rate-free}
\end{eqnarray}
We should note that $H \propto a^{-2}$ in the radiation dominated
universe and $H \propto a^{-3/2}$ in the matter dominated universe
while $\rho_{\rm t} \propto a^{-4}$ until the energy density of baryons
dominate the one of photons, which happens much later than
recombination.  Therefore $v_{\rm A} =B/\sqrt{\rho_{\rm t} + p_{\rm t}}$ is still
constant through this regime.  This time evolution is consistent with
the numerical result.

Now we can estimate the power spectrum on scales where the direct 
cascade or diffusion damping during the free streaming regime occurs.   
Even if the cascade process occurs first, the power law slope of the
decay immediately washes away by diffusion since the scales of the
direct cascade and the diffusion are very close.  Therefore we only
consider the diffusion process here.
The resultant power spectrum leads to
\begin{equation}  
{\cal P}_v(t,k)
\approx \left( \frac{(\rho_{\rm t}+p_{\rm t})k_{\rm int}}{a \alpha
    \rho_{\rm b}}  {\cal P}_{\rm A}(k_{\rm int})\right)^2 
{\rm e}^{-2( k / k_{\rm int})^2}, 
\qquad k>k_{\rm int}.
\label{power-velo-free-red-small} 
\end{equation}
In the above power spectrum, the shape on the scales larger than the
integral scale is identical to the one 
without further growth of the integral scale during free-streaming 
(cf. Eqs.~(\ref{power-velo-free-red}) - (\ref{power-velo-free-red-large})).


\subsection{Summary of Velocity Power Spectra}\label{power-summary}

Let us summarize the fluid velocity power spectrum 
of Alfv\'en modes at LSS for estimating power spectra of 
CMB anisotropies and polarization.  
There are three different cases corresponding to the integral scale 
on LSS.  We refer them as ``Case A'', ``Case B'' and ``Case C'' 
for $k_{\rm int} < k_{\rm f} = 1.2 {\rm Mpc}^{-1}$, 
$k_{\rm f} = 1.2 {\rm Mpc}^{-1} < k_{\rm int} < 8 {\rm Mpc}^{-1}$, 
and $ 8 {\rm Mpc}^{-1} < k_{\rm int} $, 
respectively
(cf. to FIG.~\ref{fig:z-vis-fs} at $z\approx 1000$). 
Here $1.2 \rm Mpc^{-1}$ and $8 \rm Mpc^{-1}$ 
correspond to amplitudes of comoving magnetic field strength
as $30\rm nGauss$ and   $16\rm nGauss$.
  Case A is the case in which  
the viscous regime continues until recombination.  
For Case B, while transition from the viscous to the 
free streaming regimes occurs before recombination, 
the direct cascade process never happens on the integral scale.   
The diffusion process on the integral scale only takes place for Case C.  

Let us first list the power spectra with $n>3/2$.  

\noindent 
\underline{Case A}   
(Eqs.~(\ref{power-velo-vis-blue-large}), (\ref{power-velo-vis-blue}) and (\ref{power-velo-vis-blue-small}))
\\
\begin{eqnarray}
{\cal P}_v(t,k)
& \approx &
\left( {\left(\rho_{\rm t}+p_{\rm t}\right) a \over \nu k_{\rm int}} 
{\cal P}_{\rm A}(t_{\rm t-v} ,k_{\rm int}) 
\right)^2
\left({k \over k_{\rm int}}\right) \propto k,
 \quad k<k_{\rm int} , \\
{\cal P}_v(t,k)
& \approx & 
\left( {\left(\rho_{\rm t}+p_{\rm t}\right) a \over \nu k} 
{\cal P}_{\rm A}(t_{\rm t-v} ,k)\right)^2 \propto k^{2(m-1)}
 , \quad k_{\rm int}<k<k_{\rm f} , \\ 
{\cal P}_v(t,k)
& \approx  &
\left( {\left(\rho_{\rm t}+p_{\rm t}\right) k \over \rho_{\rm b} \alpha  a} 
{\cal P}_{\rm A}(t_{\rm t-v} ,k)\right)^2 \propto k^{2(m+1)}
, \quad k_{\rm f}<k .
\label{power-velo-case-A-bule}
\end{eqnarray}

\noindent 
\underline{Case B}  
(Eqs.~(\ref{power-velo-free-blue-large}), (\ref{power-velo-free-blue}) 
and (\ref{power-velo-free-blue-small}))\\
\begin{eqnarray} 
{\cal P}_v(t,k)
& \approx &
\left( {\left(\rho_{\rm t}+p_{\rm t}\right) a \over \nu k_{\rm int}} 
{\cal P}_{\rm A}(t_{\rm t-v} ,k_{\rm int}) 
\right)^2
\left({k \over k_{\rm int}}\right)  
\propto k, \quad k<k_{\rm f}, \\
{\cal P}_v(t,k)
& \approx &
\left( {\left(\rho_{\rm t}+p_{\rm t}\right) k_{\rm int} \over \rho_{\rm b} \alpha a } 
{\cal P}_{\rm A}(t_{\rm t-v} ,k_{\rm int}) 
\right)^2
\left({k \over k_{\rm int}}\right)^{5}  
\propto k^5, \quad k_{\rm f}<k<k_{\rm int} ,\\
{\cal P}_v(t,k)
&\approx &
\left( {\left(\rho_{\rm t}+p_{\rm t}\right) k \over \rho_{\rm b} \alpha a } 
{\cal P}_{\rm A}(t_{\rm v-f} ,k)\right)^2 
\propto k^{2(m+1)}, \quad k_{\rm int}<k. 
\label{power-velo-case-B-blue}
\end{eqnarray}

\noindent 
\underline{Case C} 
(Eqs.~ (\ref{power-velo-free-blue-large}), (\ref{power-velo-free-blue})  
and (\ref{power-velo-free-red-small}))\\
\begin{eqnarray} 
{\cal P}_v(t,k)
& \approx &
\left( {\left(\rho_{\rm t}+p_{\rm t}\right) a \over \nu k_{\rm int}} 
{\cal P}_{\rm A}(t_{\rm t-v} ,k_{\rm int}) 
\right)^2
\left({k \over k_{\rm int}}\right)  
\propto k, \quad k<k_{\rm f}, \\
{\cal P}_v(t,k)
& \approx &
\left( {\left(\rho_{\rm t}+p_{\rm t}\right) k_{\rm int} \over \rho_{\rm b} \alpha a } 
{\cal P}_{\rm A}(t_{\rm t-v} ,k_{\rm int}) 
\right)^2
\left({k \over k_{\rm int}}\right)^{5}  
\propto k^5, \quad k_{\rm f}<k<k_{\rm int} , \\
{\cal P}_v(t,k)
&\approx & 
\left( \frac{(\rho_{\rm t}+p_{\rm t})k_{\rm int}}{a \alpha
    \rho_{\rm b}}  {\cal P}_{\rm A}(k_{\rm int}) \right)^2 
{\rm e}^{-2( k / k_{\rm int})^2},
\quad k_{\rm int}<k.
\label{power-velo-case-C-blue}
\end{eqnarray}

\begin{figure}[htbp]
  \begin{center}
    \includegraphics[keepaspectratio=true,height=60mm]{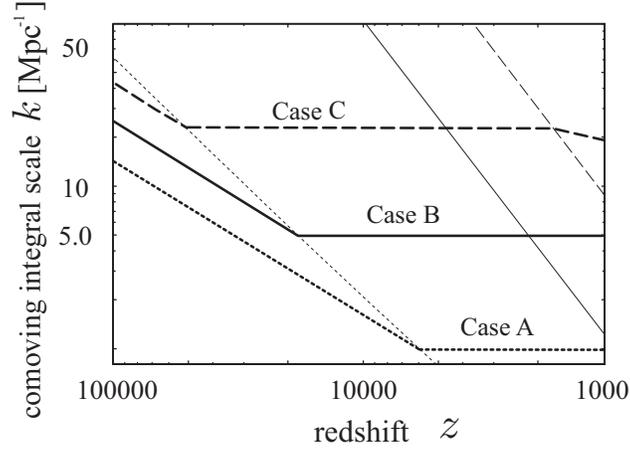}
  \end{center}
  \caption{Evolution of the comoving integral scale of the magnetic
  field configurations with $n=4$.  The dotted line corresponds to
  Case A which undergoes the turbulent regime and the viscous
  regime before recombination.  The solid line corresponds
  to Case B which passes through turbulent, viscous and free
  streaming regimes, with nevertheless no growth of the integral scale
  immediately before recombination occurring.  The dashed line
  corresponds to Case C in which there is growth of $L_{\rm int}$ in
  the free streaming regime shortly before recombination.  The thin
  dotted line represents the transition line from the turbulent regime
  to the viscous regime.  The thin solid line shows the evolution of
  $k_{\rm f}$, whereas the thin dashed line represents the start of
  growth of the integral scale in the free streaming regime.}
  \label{fig:magevolution}
\end{figure}

FIG. \ref{fig:magevolution} shows the evolutionary track of each case.
The Case A is represented as the dotted line.
The Case B and Case C are the solid and the dashed lines.
The slope of the tracks follow 
Eqs. (\ref{rate-tur}) and (\ref{rate-free}).
Here we employ the spectral index of the initial magnetic fields $n=4$.

Next we list the power spectra with $n<3/2$.  

\noindent 
\underline{Case A'} 
(Eqs.~(\ref{power-velo-vis-red}) and (\ref{power-velo-vis-blue-small}))
\\
\begin{eqnarray}
{\cal P}_v(t,k)
&\approx& \left(  {\left(\rho_{\rm t}+p_{\rm t}\right) a \over \nu k} 
{\cal P}_{\rm A}(t_{\rm t-v} ,k) \right)^2 
\propto 
\left \{
\begin{array}{ll}
k^{2(n-1)}, & \qquad k<k_{\rm int},\\
k^{2(m-1)}, & \qquad k_{\rm int}<k<k_{\rm f},
\end{array}
\right.\\
{\cal P}_v(t,k)
& \approx  &
\left( {\left(\rho_{\rm t}+p_{\rm t}\right) k \over \rho_{\rm b} \alpha  a} 
{\cal P}_{\rm A}(t_{\rm t-v} ,k)\right)^2 \propto k^{2(m+1)}, 
\quad k_{\rm f}<k.
\label{power-velo-case-A-red}
\end{eqnarray}

\noindent 
\underline{Case B'} 
(Eqs.~(\ref{power-velo-free-red-large}) and (\ref{power-velo-free-red}))\\
\begin{eqnarray}
{\cal P}_v(t,k)
& \approx & 
\left(  {\left(\rho_{\rm t}+p_{\rm t}\right) a \over \nu k} 
{\cal P}_{\rm A}(t_{\rm t-v} ,k)\right)^2 
\propto k^{2(n-1)}, \quad k<k_{\rm f} ,\\
{\cal P}_v(t,k)
&\approx & \left(  {\left(\rho_{\rm t}+p_{\rm t}\right) k \over \rho_{\rm b} \alpha a } 
{\cal P}_{\rm A}(t_{\rm v-f} ,k)\right)^2 
\propto 
\left \{
\begin{array}{ll}
k^{2(n+1)}, & \qquad k_{\rm f}<k<k_{\rm int}, \\
k^{2(m+1)}, & \qquad k_{\rm int}<k .
\end{array}
\right .
\label{power-velo-case-B-red}
\end{eqnarray}

\noindent 
\underline{Case C'} 
(Eqs.~(\ref{power-velo-free-red-large}), (\ref{power-velo-free-red}) ($k_{\rm f}<k<k_{\rm int}$) 
and (\ref{power-velo-free-red-small}))\\
\begin{eqnarray}
{\cal P}_v(t,k)
& \approx & 
\left(  {\left(\rho_{\rm t}+p_{\rm t}\right) a \over \nu k} 
{\cal P}_{\rm A}(t_{\rm t-v} ,k)\right)^2 
\propto k^{2(n-1)}, \quad k<k_{\rm f} ,\\
{\cal P}_v(t,k)
&\approx & \left(  {\left(\rho_{\rm t}+p_{\rm t}\right) k \over \rho_{\rm b} \alpha a } 
{\cal P}_{\rm A}(t_{\rm v-f} ,k)\right)^2 
\propto 
k^{2(n+1)},  \quad k_{\rm f}<k<k_{\rm int},\\
{\cal P}_v(t,k)
&\approx & \left( \frac{(\rho_{\rm t}+p_{\rm t})k_{\rm int}}{a \alpha
    \rho_{\rm b}}  {\cal P}_{\rm A}(k_{\rm int} )\right)^2 
{\rm e}^{-2( k / k_{\rm int})^2},
\quad k_{\rm int}<k.
\label{power-velo-case-C-red}
\end{eqnarray}

Note that on large scales ($k< k_{\rm int}$), the power law slopes 
of above equations may be steeper as is pointed out before.
In the most extreme case, ${\cal{P}}_v(t,k) \propto k^5$ instead of 
$k$ for $n>3/2$, and ${\cal P}_v(t,k) \propto k^{2(n+1)}$   instead of 
$k^{2(n-1)}$ for $n<3/2$ as is expected from the linear analysis.

\section{Calculation of  
CMB temperature anisotropy and polarization spectra}\label{seccalcmb}

Alfv\'en modes of baryon-electron fluid produced by the primordial
magnetic fields generate CMB temperature fluctuations by a Doppler
shift \cite{s-b-temp}.  Moreover, the quadrupole component of the
generated temperature anisotropies produces polarization due to
Thomson scattering. It is known that CMB polarization can be
decomposed into two parity independent modes, i.e., the E-mode
(electric type) and the B-mode (magnetic type) \cite{s-z}.  Among
them, the B-mode polarization is not primarily produced by the scalar
type perturbations which eventually form the structure of the universe
and provide dominant contribution on the temperature anisotropies.
Therefore the B-mode polarization possibly can be a good probe of 
primordial magnetic fields.

First, we summarize {the derivations of the temperature anisotropy} and
polarization spectra induced by Alfv\'en modes with use of the linear
perturbation theory \cite{s-b-temp, s-b-poral, m-k-k}.  The Alfv\'en
mode in the linear MHD theory is identical to the vector mode of the
cosmological fluid velocity in the cosmological perturbations.  The
vector metric perturbations in the linear perturbation theory can be
written as (cf. \cite{k-s-pert} also for notation)
\begin{equation}
\delta g_{0i} =-a^2 V_i =-a^2 V Q_i,\qquad \delta g_{ij} =2 a^2 H_{Tij} = 2a^2 H_T Q_{ij},
\label{metric}
\end{equation}
where $V$ and $H_T$ are the amplitudes of the vector mertric perturbations and 
$Q_i$ and $Q_{ij}$ are the vector type mode functions which are defined as
\begin{equation}
\nabla^2 Q_i=-k^2Q_i,\qquad \nabla^i Q_i=0,
\qquad Q_{ij}=-\frac{1}{2k}(\nabla _i Q_j + \nabla _j Q_i).
\label{vector-def}
\end{equation}
The amplitude $v_{\rm b}$ of the vector components of the baryon fluid
velocity $v_{{\rm b}i}$ can be represented with the mode functions as
\begin{equation}
v_{{\rm b}i} = v_{\rm b} Q_i.
\label{vector-deco}
\end{equation}
Employing the total angular momentum method \cite{h-w}, we obtain the
CMB temperature anisotropies $\Delta T ^{TT} (l)$ induced by Alfv\'en modes as
\begin{equation}
\Delta T ^{TT} (l) =T_0 \sqrt{ \frac{l (l+1) C^{TT}(l)}{2 \pi}} ,
\label{tempani}
\end{equation}
\begin{equation}
\Delta T ^{BB} (l) =T_0 \sqrt{ \frac{l (l+1) C^{BB}(l)}{2 \pi}} ,
\label{bbani}
\end{equation}
\begin{equation}
C^{TT}(l) =\frac{4}{\pi}
\int_{0}^{\infty }{k^{2}dk}\quad \langle |{\Theta^{(1)} _l (\eta_0,k) \over 2l+1}|^{2} \rangle, 
\label{ttcl}
\end{equation}
\begin{equation}
C^{BB}(l) =\frac{4}{\pi}
\int_{0}^{\infty }{k^{2}dk} \langle |{B^{(1)} _l (\eta_0,k) \over 2l+1}|^{2} \rangle, 
\label{bbcl}
\end{equation}
\begin{eqnarray}
&&{\Theta^{(1)}_l (k,\eta ) \over 2 l+1 } =  \int^{\eta} _{0} d \eta'
e^{-\tau (\eta ') } \left[\dot \tau(v_{\rm b} -V) j^{(11)} _l (k(\eta-\eta')) \right.
\nonumber \\ 
&&\qquad \qquad \qquad
\left. + \left( \dot \tau P^{(1)}(k, \eta') + {1 \over \sqrt{3} } k V \right) j^{(21)} _l (k(\eta-\eta'))\right],
\label{thetal}
\end{eqnarray}
\begin{equation}
{B^{(1)}_l (k,\eta ) \over 2 l+1 } = - {\sqrt{6} } 
\int^{\eta} _{0} d \eta'
e^{-\tau (\eta ')} \dot \tau  P^{(1)}(k, \eta') \beta  ^{(1)} _l (k(\eta-\eta')),
\end{equation}
\begin{equation}
P^{(1)} {(k,\eta)}={1 \over 10} [\Theta^{(1)} _2 (k,\eta)-\sqrt{6} E^{(1)}_2(k,\eta)],
\label{sourceterm}
\end{equation}
\begin{equation}
{E^{(1)}_2 (k,\eta ) } = -5 {\sqrt{6} } \int^{\eta} _{0} d \eta'
e^{-\tau (\eta ')} \dot \tau P^{(1)}(k, \eta') \varepsilon   ^{(1)} _2 (k(\eta-\eta')),
\end{equation}
\begin{equation}
j^{(11)} _l (x)=\sqrt{l (l+1) \over 2} {j_l (x) \over x},  
\quad j^{(21)} _l (x)=\sqrt{{3l (l+1)\over 2}} {d \over dx} \left({j_l(x)\over x}\right),
\end{equation}
\begin{equation}
\varepsilon ^{(1)} _2 (x)={j_2(x) \over x^2}+{1 \over x}{d {j_2(x)} \over dx} ,
\end{equation}
\begin{equation}
\beta ^{(1)} _l (x)={1 \over 2} \sqrt{(l-1)(l+2)}{j_l(x) \over x},
\end{equation}
\begin{equation}
e^{-\tau (\eta ')}=  \exp \left( - \int^{\eta} _{\eta'} \dot \tau d \eta'' \right),
\label{opti}
\end{equation}
where $j_l(x)$ is the spherical Bessel function, $\eta \equiv \int
da/a $ is the conformal time, 
subscript $0$ denotes the present epoch and $\dot
\tau$ is the differential optical
depth which is expressed as
\begin{equation}
\dot \tau = a n_e \sigma_{\rm T} = a /L_{\rm mfp}.
\label{taudott}
\end{equation}
For further notation the reader is referred to Ref.~\cite{h-w}.
The conformal time $\eta$ can be written in terms of
redshift $z$ as 
\begin{equation}
\eta={1 \over H_0} \int 
{dz \over \sqrt{  1-\Omega_{\rm M} + \Omega_{\rm M} [1+(1+z)(1+z_{\rm eq}^{-1})](1+z)^3}},
\end{equation}
where $z_{\rm eq} = 2.4 \times 10^4 \Omega_{\rm M} h^2$ is the redshift of
the matter-radiation equality epoch.  
In the above expressions, we have taken two independent vector modes
into account.  The time evolution of $v_{\rm b}$ is given by the Euler
equation.  


Combining these formulas with the linear perturbation theory, we can
calculate temperature anisotropy and polarization spectra induced by
the nonlinear Alfv\'en modes by simply substituting the fluid velocity
of Alfv\'en modes $v$ in Sec. \ref{nonlinear evolution} to $v_{\rm
b}-V$ of Eq. (\ref{thetal}) due to the Newtonian gauge condition
\cite{h-w}.  Although, we obtain the Alfv\'en modes from (fully
non-linear) numerical simulation, we still can treat them as the vector
modes because $\nabla \cdot {\bf v} =0$.

For the recombination history, we assume the visibility function 
to be a Guassian as 
\begin{equation}
g(\eta ) \equiv
\dot \tau \exp \left( - \int^{\eta_0} _{\eta} \dot \tau d \eta'
\right)= (2 \pi \sigma^2)^{-1/2} \exp \left[ -{(\eta-\eta_{\rm LSS})^2 \over(2 \sigma^2)} \right], 
\label{visi-gauss}
\end{equation}
where $\eta_{\rm LSS}$ is the conformal time at the LSS and $\sigma$
is the width of the LSS.  From the WMAP results, the 
redshift and the thickness of the LSS are $1+z=1089$ and $\Delta
z=195$ \cite{wmap}.  This implies that $\sigma$ is 13Mpc for the
cosmological parameters we use in this work.

In our calculations the magnetic field power spectrum stays 
virtually unchanged after the transition from turbulence to viscous diffusion (i.e. for $t > t_{t-v}$),
even in the case where $k_{\rm int}= 10 {\rm Mpc^{-1}}$ (Case C), since
the free streaming regime before recombination is so short that there is hardly any time
for $L_{\rm int}$ to grow.  Since in this study we concentrate on causal spectra ($n>3/2$) 
we mostly require the magnetic filed  spectrum around the integral scale in 
order to derive the fluid velocity spectrum on all scales. 
We therefore take the numerical results in
the fully developed turbulent regime at  $s=2.49$, which is expected to resemble that at $t_{t-v}$ closely.
From these numerical spectra of the magnetic fields, we obtain the
power spectra of Alfv\'en modes using
Eqs. (\ref{power-velo-case-A-bule})--(\ref{power-velo-case-C-red}).
We assume the equipartition between magnetic fields and fluid
velocities during the turbulent regime, which makes the power spectra
flat on large scales (small $l$'s).  The modification of the power
spectra due to possible violation of this
assumption will be discussed in the next section.  

For the calculation of polarization, we compute the source term
$P^{(1)}$ by using the publicly available code CAMB \cite{al, camb} in
which we substitute Alfv\'en modes obtained by the numerical
simulation.

\begin{figure}[tbp]
  \begin{center}
    \includegraphics[keepaspectratio=true,height=65mm]{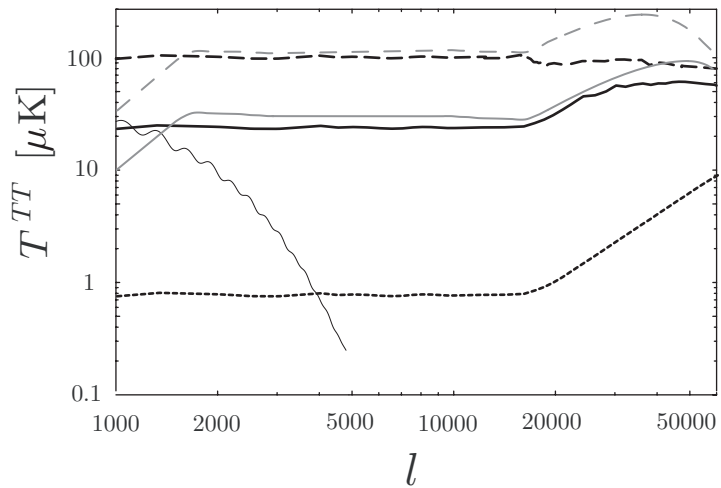}
  \end{center}
  \caption{
  The temperature anisotropy power spectrum.  
  The dashed line is the case of the magnetic fields with the comoving integral scale $k_{\rm int}=1.0 {\rm  Mpc}^{-1}$ with 32nGauss. 
  The solid line is the case of $k_{\rm int}=1.5 {\rm  Mpc}^{-1}$
  with 28nGauss
  and the dotted line is the case of $k_{\rm int}=10 {\rm  Mpc}^{-1}$ with 16nGauss.
  The comoving integral scale is related with the magnetic field strength by Eq. (\ref{eq:int_viscus}).
  The gray line is the linear result of the magnetic field strength $28$nGauss at 1.5Mpc with 
  spectral index $n=4$.
  This line has exponential damping at smaller scales than the cut-off scale (for details see Sec.~\ref{compare:linear}).
  For purpose of comparison, we plot the temperature anisotropy power spectrum 
  in the standard $\lambda$CDM model (a thin line), 
  computed using CAMB with the same cosmological parameters \cite{camb}.
  }
  \label{fig:temp-ani}
\end{figure}

\begin{figure}[tbp]
  \begin{center}
    \includegraphics[keepaspectratio=true,height=65mm]{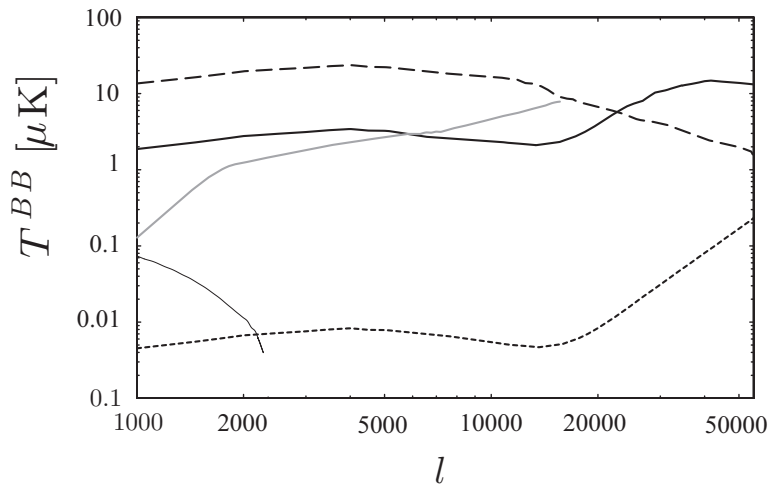}
  \end{center}
  \caption{
  The B-mode polarization power spectrum.  
  The dashed line is the case of the magnetic fields with the comoving integral scale $k_{\rm int}=1.0 {\rm  Mpc}^{-1}$ with 32nGauss
  The solid line is the case of $k_{\rm int}=1.5 {\rm  Mpc}^{-1}$
  with 28nGauss and the dotted line is the case of $k_{\rm int}=10 {\rm  Mpc}^{-1}$ with 16nGauss.
  The comoving integral scale is related to the magnetic field strength by Eq. (\ref{eq:int_viscus}).
  The gray line is the linear result of the magnetic field strength $28$nGauss at 1.5Mpc with 
  spectral index $n=4$.
  Because this linear result uses the tight coupling approximation,
  it is plotted up to the $\sim k_f\approx 1.4\,$Mpc$^{-1}$.
  For purpose of comparison, we plot the polarization power spectrum 
  in the standard $\lambda$CDM model (a thin line), computed using CAMB with the same cosmological parameters. 
  This anisotropies is produced by the gravitational lensing mainly.}
  \label{fig:poral-ani}
\end{figure}

Now we are ready to calculate CMB temperature anisotropy and
polarization spectra.  Hereafter, we fix the initial power law index
$n$ to be $4$.
Temperature anisotropy and B-mode polarization spectra
are shown in FIGs.~\ref{fig:temp-ani} and \ref{fig:poral-ani},
respectively.  Here we calculate models with $k_{\rm int}= 10 {\rm
Mpc^{-1}}$, $1.5 {\rm Mpc^{-1}}$ and $1.0 {\rm Mpc^{-1} }$ at the LSS,
whose magnetic field strengths at the integral scale are $15$nGauss,
$28$nGauss, and $32$nGauss, respectively (see Eq.  (\ref{eq:int_viscus})).
Note that $k_{\rm int}= 10 {\rm Mpc^{-1}}$ corresponds to Case C, $1.5
{\rm Mpc^{-1}}$ to Case B and $1.0 {\rm Mpc^{-1} }$ to Case A in
Sec.~\ref{power-summary}.

For purpose of comparison, we also plot the temperature
anisotropies and polarization by using linear perturbations of the
Alfv\'en modes in FIGs.~\ref{fig:temp-ani} and \ref{fig:poral-ani}.  Here
we adopt two types of the magnetic field strength.
One is $32$nGauss at $1.0{\rm Mpc^{-1}}$ and another is $28$nGauss at $1.5{\rm
  Mpc^{-1}}$.
In both cases we take the spectral index as $n=4$.

\section{Discussion}

Let us investigate the features of the power spectra of CMB 
anisotropies and polarization obtained in the previous section.  
For the qualitative understanding of  these spectra, 
we first develop analytic expressions of CMB anisotropies and
polarization.
 
Both spectra are induced from the power spectrum of the 
fluid velocity $P_v(k)$ at the LSS.  In fact, employing 
the small angle approximation and considering the phase cancellation
damping of anisotropies within the thickness of the LSS, 
we obtain \cite{s-b-temp, s-b-poral}
\begin{equation}
\Delta T^{TT}(l) 
\approx T_0 \left.  \sqrt{{\sqrt{\pi} \over 2 k \sigma} {\cal P} _v(k )} 
 \right |_{k=l/(\eta_0-\eta_{\rm LSS})},
\label{temp-ani-velocity}
\end{equation}
\begin{equation}
\Delta T^{BB}(l) 
\approx T_0 \left . 
\sqrt{ {3\sqrt{\pi} \over 2 k \sigma}  {\cal P}_{P^{(1)}}(k )} 
\right |_{k=l/(\eta_0-\eta_{\rm LSS})},
\label{poral-ani-velocity}
\end{equation}
where ${\cal P}_{P^{(1)}}(k )$ is the power spectrum of the source term
defined as 
\begin{equation}
{\cal P}_{P^{(1)}} (k )\equiv 
{k^3 \over 2 \pi^2} \langle |P^{(1)}|^2 \rangle .
\end{equation}
Here, $1/k \sigma$ is due to the phase cancellation 
damping within the thickness of the LSS.  

In Sec. \ref{power-spectrum-alfven} we studied the velocity power
spectra where we found that the shape and amplitude of the power
spectra on the LSS depend on the regime of the comoving integral scale at
recombination. Note that magnetic fields with the comoving integral scale
$k_{\rm int} =1.2{\rm Mpc}^{-1}$ at recombination are transiting from
the viscous regime to the free streaming regime right at recombination
as is shown in
FIG.~\ref{fig:magevolution}.  At the recombination epoch, the integral
scale is still in the viscous regime for $k_{\rm int} < 1.2{\rm
Mpc}^{-1}$, while it has already been in the free streaming regime for
$k_{\rm int} > 1.2{\rm Mpc}^{-1}$.

In the viscous or free streaming regimes, the velocity power spectra
do not explicitly depend on the initial spectral index of the magnetic
fields $n$ if $n>3/2$, which is the case we consider in this paper, as
is shown in \ref{velo-viscoussection} and \ref{velo-fssection} (or
summarized in \ref{power-summary}).  Even in this initially blue
spectrum case, however, the initial spectral index still affects the
growth rate of the integral scale (see Eqs. (\ref{rate-tur}) for the
turbulent regime and (\ref{rate-free}) for the free streaming regime).
But as is shown in FIG.~\ref{fig:magevolution}, the integral scale
does not change during the viscous regime and only changes little in
the free streaming regime, provided the magnetic fields are
sufficiently strong, i.e. the comoving $B > 16\,$nGauss at
recombination.  Since the turbulent regime, where the integral scale
mostly evolves, is long before the recombination epoch, we can ignore
the growth of the integral scale and accordingly the dependence of the
initial spectral index.  Therefore we expect the velocity power
spectra and the resultant temperature anisotropy and polarization
spectra to show virtually no dependence on the initial spectral index
$n$, for $n > 3/2$.

\subsection{Temperature anisotropy power spectrum}

Let us now discuss the temperature anisotropies.  Following
Sec.~\ref{power-summary}, we investigate the temperature anisotropy
spectrum for three separate cases corresponding to the integral scale
on the LSS.

First, we consider Case A in which the integral scale $k_{\rm int}
< 1.2 {\rm Mpc}^{-1}$.  The integral scale is still in the viscous
regime at the recombination epoch in this case.  
Substituting the equations of Case A in Sec. \ref{power-summary}
into Eq. (\ref{temp-ani-velocity}),
we can derive the approximations of the temperature anisotropy spectrum
produced by the magnetic fields as 
\begin{eqnarray}
\Delta T^{TT} &&\approx 
T_0   \left.
\sqrt{
{\sqrt{\pi} \over 2 k \sigma} 
\left( { \left(\rho_{\rm t}+p_{\rm t}\right) a \over \nu k_{\rm int}} 
{\cal P}_{\rm A}(t_{\rm t-v} ,k_{\rm int}) 
\right)^2
\left({k \over k_{\rm int}}\right)
}\right |_{k=l/(\eta_0-\eta_{\rm LSS})}
\nonumber \\
&&\approx
104 \left(\frac{k_{\rm int}}{1.0 {\rm Mpc}^{-1}} \right)^{-13/6}~[\mu {\rm K}], 
\qquad l<l_{\rm int},
\label{temp-ani-vis-large}
\end{eqnarray}
\begin{eqnarray}
\Delta T^{TT} &&\approx
T_0 \left. 
\sqrt{
{\sqrt{\pi} \over 2 k \sigma} 
\left( {\left(\rho_{\rm t}+p_{\rm t}\right) a \over \nu k} 
{\cal P}_{\rm A}(t_{\rm t-v} ,k)\right)^2
} 
\right |_{k=l/(\eta_0-\eta_{\rm LSS})}
\nonumber \\
&&\approx
104 \left( 
\frac{k_{\rm int}}{1.0 {\rm Mpc}^{-1}} \right)^{-13/6}  
\left( \frac{l}{l_{\rm int}} \right)^{m-3 /2}~[\mu {\rm K}],
\qquad l_{\rm int}<l<l_{\rm f},
\label{temp-ani-vis-small}
\end{eqnarray}
\begin{eqnarray}
\Delta T^{TT} 
&&\approx T_0 \left.  
\sqrt{
{\sqrt{\pi} \over 2 k \sigma}
\left( {\left(\rho_{\rm t}+p_{\rm t}\right) k \over \rho_{\rm b} \alpha  a} 
{\cal P}_{\rm A}(t_{\rm t-v} ,k)\right)^2
 }
\right |_{k=l/(\eta_0-\eta_{\rm LSS})}
\nonumber \\
&&\approx 31.5 \left(\frac{k_{\rm int}}{1.0{\rm Mpc}^{-1}} \right)^{-1/6}  
\left( \frac{l}{l_{\rm int}} \right)^{m+1 /2}~[\mu {\rm K}],
\qquad l_{\rm f}<l,
\label{temp-ani-vis-very-small-nodirect}
\end{eqnarray}
where 
\begin{equation}
l_{\rm int}=k_{\rm int} (\eta_0 -\eta_{LSS})
=14500 \left( \frac{k_{\rm int}}{1.0 {\rm Mpc}^{-1}}\right),
\end{equation}
\begin{equation}
l_{\rm f}=k_{\rm f} (\eta_0 -\eta_{LSS})
=17800.
\end{equation}

We find a very good agreement between these approximations and 
full numerical calculations as is shown in
FIG. \ref{fig:temp-ani}. 
On scales $l < l_{\rm int}$ and for $k\sigma > 1$,
Eq.~(\ref{temp-ani-vis-large}) suggests no $l$ dependence of the
temperature anisotropy spectrum, whereas for $l < l_{\rm int}$ and 
$k\sigma < 1$, although this region is not shown in FIG. \ref{fig:temp-ani}, 
a residual small $l$-dependence is expected to remain 
$\propto l^{1/2}$.  
Since $m \simeq -2/3$, $l$
dependence on scales $l > l_{\rm f}$ is also very weak.  We expect to
see some variation of the temperature anisotropy spectrum between $l_{\rm
int} < l < l_{\rm f}$.  The dashed line in FIG. \ref{fig:temp-ani}
shows these features.  

Note that we assume the equipartition between magnetic fields and
fluid velocities during the turbulent regime.  On very large scales
(small $l$'s), we may expect violation of this assumption.  Above the
scale at which this assumption is no longer valid, the temperature
anisotropy spectrum shows $l$ dependence.  In the most extreme case,
the advective term can be completely ignored.  Therefore the behavior
of the power spectrum is described by the linear perturbation as is
shown in Appendix (A29), and the temperature anisotropy spectrum is
proportional to $l^{5/2}$.  In FIG.~4, this damping on small $l$'s
(the blue spectrum) is shown for the linear case.  For the nonlinear
cases, we do not really know the location of the break from the flat
spectrum and the power law index since they strongly depend on when
and how the initial magnetic fields were formed.  If there existed strong
magnetic fields in the very early universe, the assumption of the
equipartition is mostly valid, and the flat spectrum on small $l$'s is
expected.

The amplitude of temperature anisotropies in Case A is generally 
very large.  Even in the case when $k_{\rm int}=1.2 {\rm Mpc}^{-1}$,
which corresponds to the smallest amplitude of Case A (or transition 
between Case A and Case B),
$\Delta T^{TT} \simeq 70 \mu \rm K$.  This
amplitude is much larger than the one measured by WMAP or BOOMERANG,
i.e., $\Delta T^{TT} \simeq 40 \mu \rm K$ at $l \simeq 800$
\cite{bennett} or $\Delta T^{TT} \simeq 30 \mu \rm K$ at $l \simeq
1100$ \cite{BOOM_TMP}, in which major parts of anisotropies should be
explained as the primordial anisotropies.  Only a small fraction of
anisotropies can be interpreted as primordial magnetic field origin at
most.  
Note that we ignore here the possible damping of the
temperature spectrum on small $l$'s due to the non-equipartition
between magnetic fields and fluid velocities during the turbulent
regime. Accordingly the amplitude of temperature anisotropies at 
$l=800$ and $1100$ may be smaller than $70 \mu \rm K$.  
Even in the most extreme case with assuming the same power spectrum as
the linear calculation, however, we still get $\Delta T^{TT} \simeq 20
 \mu \rm K$ at $l=1100$, which is about $3/2$ of the observed
temperature anisotropies.  Here we take the break scale $l_{\rm v}$ is
$1800$ using the expression in Appendix, and employ the equation 
$\Delta T^{TT} = 70 (l/l_{\rm v})^{5/2} \mu \rm K$. 

\begin{figure}[htbp]
  \begin{center}
    \includegraphics[keepaspectratio=true,height=70mm]{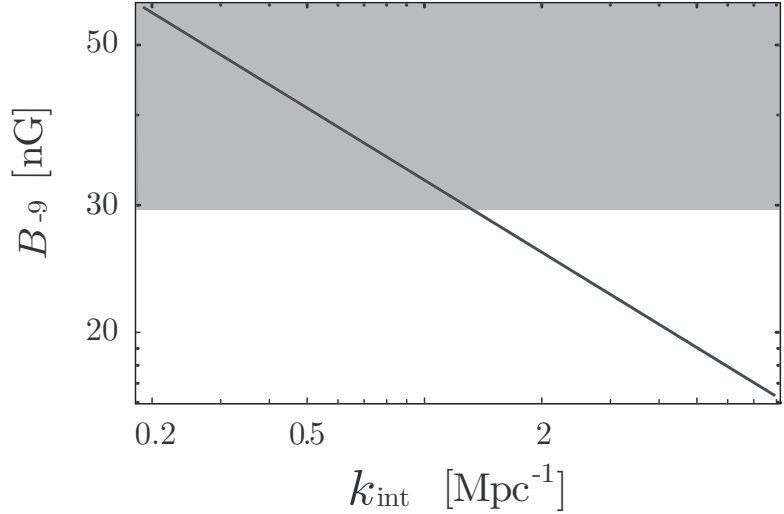}
  \end{center}
  \caption{
  The magnetic field strength $B_{-9}$ for the comoving integral
  scale $k_{\rm int}$ (Eq.~(\ref{eq:int_viscus})).  
  The gray region is the forbidden region by the WMAP results.
  Magnetic fields with comoving strength larger than $30$nGauss 
  at recombination are ruled out.}
  \label{fig:magallowed}
\end{figure}

Therefore we can safely rule out the possibility of  Case A, i.e., 
$k_{\rm int} < 1.2 {\rm Mpc}^{-1}$ and $B > 30$nGauss at the recombination
epoch (see FIG. \ref{fig:magallowed}) for $n > 3/2$.

Now we study Case B. In this case, the integral scale is $1.2 {\rm
Mpc}^{-1} \lesssim k_{\rm int} \lesssim 8 {\rm Mpc}^{-1}$ and 
the temperature anisotropies on the LSS are given by the power spectrum in
the free streaming regime.  Substituting the equations of Case B
into Eq. (\ref{temp-ani-velocity}), we get
\begin{eqnarray}
\Delta T^{TT} &&\approx 
T_0 \left.  
\sqrt{
{\sqrt{\pi} \over 2 k \sigma} 
\left[ \frac{1}{\nu} \frac{a}{k_{\rm int}} {\cal P}_{\rm A } (k) 
\left( \frac{k}{k_{\rm int}} \right)^{-n} \right]^2 \left({ k\over k_{\rm int} }\right)
} 
\right |_{k=l/(\eta_0-\eta_{\rm LSS})}
\nonumber \\
&& \approx 43 \left(\frac{k_{\rm int}}{1.5 {\rm Mpc}^{-1}} \right)^{-13/6}
~[\mu {\rm K}],  
\qquad l<l_{\rm f},
\label{temp-fs-very-large}
\end{eqnarray}
\begin{eqnarray}
\Delta T^{TT} &&\approx 
T_0 \left.  
\sqrt{
{\sqrt{\pi} \over 2 k \sigma}
\left( {\left(\rho_{\rm t}+p_{\rm t}\right) k_{\rm int} \over \rho_{\rm b} \alpha a } 
{\cal P}_{\rm A}(t_{\rm t-v} ,k_{\rm int}) 
\right)^2
\left({k \over k_{\rm int}}\right)^{5}
}
\right |_{k=l/(\eta_0-\eta_{\rm LSS})}
\nonumber \\
&& \approx 56 \left(\frac{k_{\rm int}}{1.5 {\rm Mpc}^{-1}} \right)^{-1/6} 
 \left( \frac{l}{l_{\rm int}} \right)^{2}~[\mu {\rm K}],
\qquad l_{\rm f}<l<l_{\rm int},
\label{temp-fs-large}
\end{eqnarray}
\begin{eqnarray}
\Delta T^{TT} 
&&\approx 
T_0 \left.  
\sqrt{
{\sqrt{\pi} \over 2 k \sigma} 
\left( {\left(\rho_{\rm t}+p_{\rm t}\right) k \over \rho_{\rm b} \alpha a } 
{\cal P}_{\rm A}(t_{\rm v-f} ,k)\right)^2 
} 
\right |_{k=l/(\eta_0-\eta_{\rm LSS})}
\nonumber \\
&&\approx 56 \left(\frac{k_{\rm int}}{1.5 {\rm Mpc}^{-1}} \right)^{-1/6} 
 \left( \frac{l}{l_{\rm int}} \right)^{m+1/2}~[\mu {\rm K}],
\qquad l_{\rm int}<l.
\label{temp-fs-small-nodirect}
\end{eqnarray}
 
Again we find a good agreement between the above approximations and
full numerical calculations, i.e., a solid line in
FIG. \ref{fig:temp-ani}.  On the scales $l < l_{\rm f}$ and $l > l_{\rm
int}$, we expect no, or very little $l$ dependence from
Eqs. (\ref{temp-fs-very-large}) and (\ref{temp-fs-small-nodirect})
while the non-equipartition between magnetic fields and fluid
velocities may change $l$ dependence in the same manner as Case A.
Between $l_{\rm f} < l < l_{\rm int}$, however, temperature fluctuations
should increase as $l^2$.  In FIG. \ref{fig:temp-ani}, the solid
line stays flat on the scale $l < 18000 \simeq l_{\rm f}$, and increases
until $l \simeq 30000$.  However $\Delta T^{TT} \propto l^2$ only on 
the scale $l \lesssim 22000 \simeq l_{\rm int}$ and gradually decreases
the gradient for larger $l$.

The WMAP and BOOMERANG results have ruled out Case A.  They also constrain
Case B as $k_{\rm int} > 1.5 {\rm Mpc}^{-1}$, or $B < 29 \rm nGauss$
at the recombination epoch if the equipartition assumption is valid.  
When the assumption is violated, we cannot set concrete limits for
Case B.
On the other hand, one might think that
CMB anisotropies produced by the magnetic fields can explain the small
scale excess of the temperature power spectrum observed by the
CBI experiment \cite{cbi}.  
However, the power spectrum of CMB
anisotropies we find is very flat on scales $ 2000 < l < 18000$ even
with considering possible violation of the equipartition.  Therefore it
is rather difficult to explain the CBI experiment which shows Silk
damping at $l=2000$ and increases the power to  $l=2800$.

Finally let us investigate Case C, 
in which some further growth of the integral scale
takes place before recombination.
In this case, the power
spectrum Eq. (\ref{temp-fs-small-nodirect}) is no longer valid due to
the direct cascade while the rests of the power spectra
Eq. (\ref{temp-fs-very-large}) and (\ref{temp-fs-large}) are the same.
Therefore the difference between Case B and Case C appears in
the power spectrum at scales smaller than the integral scale.
Employing Eq. (\ref{power-velo-case-C-blue}), we obtain the temperature
anisotropy spectrum as
\begin{eqnarray}
\Delta T^{TT} &&\approx T_0 \left.  \sqrt{ {\sqrt{\pi} \over 2 k
\sigma} \left[ \frac{(\rho_{\rm t}+p_{\rm t})k_{\rm int}}{a \alpha
\rho_{\rm b}} \right]^2 {\cal P}_{\rm A}(k_{\rm int})^2
{\rm e}^{-2\left(k/{k_{\rm int}}\right)^{2} } }
\right|_{k=l/(\eta_0-\eta_{\rm LSS})} \nonumber \\ &&\approx 43
\left(\frac{k_{\rm int}}{10 {\rm Mpc}^{-1}} \right)^{-1/6} 
{\rm e}^{-\left(l/l_{\rm int}\right)^2}~[\mu {\rm K}], 
\qquad l_{\rm int}<l.
\label{temp-fs-small}
\end{eqnarray}

The dotted line of FIG \ref{fig:temp-ani} corresponds to Case C. In
this figure, we can find the temperature anisotropy spectrum stays
constant for $l < l_{\rm f}$ and follows $\propto l^2$ for $l> l_{\rm f}$.
The integral scale of this model is too large, i.e., $l_{\rm int} =
1.5 \times 10^{5}$, to see the difference with Case B.

Note that the temperature anisotropy spectrum induced by 
the magnetic fields shown in  
FIG. \ref{fig:temp-ani}  does not show any significant damping on large 
$l$'s (small scales) unlike the primary temperature anisotropy spectrum which 
suffers Silk damping at $l \simeq 2000$.   
This is due to the fact that
Alfv\'en modes can survive even below the Silk damping scale~\cite{j-k-o}.
The damping scale for Alfv\'en modes is the integral scale.  
Our numerical simulations also show the damping below the integral scale is 
rather mild compared to the linear calculations.
There exists the
diffusion scale $v_{\rm A}/k_{\rm S}$ below which the power spectrum
is exponentially damped for Case C (and Cases A and B in extremely
small scales), while we cannot see the damping
within the range of 
FIG. \ref{fig:temp-ani}. 
Further comparison
with linear calculations will be made in Sec. \ref{compare:linear}.

\subsection{B-mode polarization power spectrum}

We investigate the B-mode polarization spectrum shown in
FIG.~\ref{fig:poral-ani}.  For the quantitative understanding, we need
to know the behavior of the source term $P^{(1)}$ defined in 
Eq. (\ref{sourceterm}).  In our calculation, we employ CAMB to solve
$P^{(1)}$.  However, it is rather easy to develop an analytic
solution for the evolution of $P^{(1)}$.

First of all, we divide the evolution into two stages.  The first one
is the period when the tight coupling approximation is valid.  Once
the wave length of the perturbations becomes shorter than the Silk
scale, however, electrons and photons are decoupled and the tight
coupling approximation is no longer valid.  This gives rise to the
second stage.

In the first stage, the solution of the Boltzmann equations of
polarization becomes $E^{(1)}_2=-\left(\sqrt{6}/4\right)
\Theta^{(1)}_2 $ in the limit of $\tau \rightarrow \infty$ (Eq. (89)
of \cite{h-w}).  The quadrupole ($l=2$) component of the Boltzmann
temperature hierarchy gives $\Theta^{(1)}_2 =\left(4\sqrt{3}/9\right)
\left( k/{\dot \tau} \right)\Theta^{(1)}_1$.  Due to the tight
coupling, electron (baryon) velocity follows the photon velocity as
$\Theta^{(1)}_1 =v$.  Therefore the source term can be written as
(Eq. (94) of \cite{h-w})
\begin{equation}
P^{(1)} \equiv {1\over 10 } [\Theta^{(1)}_2-\sqrt{6}E^{(1)}_2]
= {1 \over 4 } \Theta^{(1)}_2 = 
{\sqrt{3} \over 9 } { k \over {\dot \tau}}\Theta^{(1)}_1= 
{\sqrt{3} \over 9}{k L_{\rm mfp} \over  a} v.
\label{tight_approx}
\end{equation}

In the second stage, the E-mode component is damped due to diffusion.
Therefore $\Theta^{(1)}_2$ induced by the Alfv\'en modes
(Eq. (\ref{thetal})) $v$ only contributes to the source term $P^{(1)}$
in Eq. (\ref{sourceterm}).  Namely, $P^{(1)}$ is induced by $v$.  
It is known that the oscillations of the temperature quadrupole, i.e.,
$j^{(11)}_2\left(k(\eta-\eta)\right)$ of Eq. (\ref{thetal}), suffers
damping due to the phase cancellation within the optical depth.   
Note that the $1/\sqrt{k \sigma}$ coefficient of
Eqs. (\ref{temp-ani-velocity}) and (\ref{poral-ani-velocity}) appeared
because of this effect.  Therefore we expect to have a factor
$1/\sqrt{k L_{\rm mfp}/a}$ in the source term $P^{(1)}$.  Adding a
numerical factor to fit with the simulation, we obtain
\begin{equation}
P^{(1)}\approx {v \over 10  \sqrt{k L_{\rm mfp}/a}}, 
\label{detight_approx}
\end{equation}
which we refer as the decoupling approximation.

\begin{figure}[htbp]
  \begin{center}
    \includegraphics[keepaspectratio=true,height=50mm]{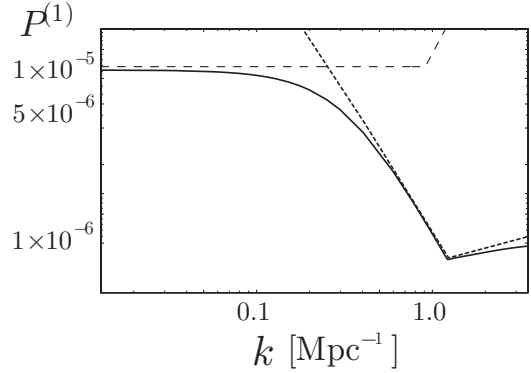}
  \end{center}
  \caption{The Thomson scattering source term $P_1(k)$ at the
    recombination epoch obtained by CAMB for a fixed magnetic field
    with 1nGauss amplitude and the scale-invariant power spectrum (solid
    line).   
  The dashed and dotted lines  are  
  the tight coupling approximation, Eq. (\ref{tight_approx}), and 
  the decoupling approximation, Eq. (\ref{detight_approx}),
    respectively.  
    Note that the Silk damping scale here is $k_{\rm S} \simeq 0.08 \rm Mpc^{-1}$, 
    which corresponds to the transition scale between tight coupling
    and decoupling and the free streaming scale is $k_{\rm f} \simeq 1.2 \rm Mpc^{-1}$.
}
  \label{fig:source}
\end{figure}

For the intuitive understanding, we plot the source term
$P^{(1)}(k,\eta_{\rm LSS})$ of the model in which the magnetic fields have
a scale-invariant power spectrum, i.e., $k^3 P_v(k) \propto k^0$, with
a magnetic field strength of nGauss in FIG. \ref{fig:source} by employing
CAMB.  In this figure, it is found that the numerical calculation
follows the tight coupling approximation at scales larger than the
Silk damping scale, $k_{\rm S} = a\sqrt{H/L_{\rm mfp}} \simeq 0.08 \rm
Mpc^{-1}$.  It is also shown that the numerical calculation traces the
decoupling approximation Eq.~(\ref{detight_approx}) on scales smaller
than the Silk damping scale.  Gradual increase of the source term on
scales $k > k_{\rm f} \equiv \sqrt{5}a/L_{\rm mfp} \simeq 1.2 \rm
Mpc^{-1}$ is due to the increase of the velocity $v$ since the
viscosity is no longer efficient and the fluid velocity can evolve on
these scales (see Eq. (\ref{fs-velo})).

Now we are ready to calculate the B-mode polarization spectrum 
using Eq. ({\ref{poral-ani-velocity}}), since the source term $P^{(1)}$ can be written by
employing the fluid velocity $v$ as Eqs. (\ref{tight_approx}) and
(\ref{detight_approx}). 
Following our calculations of temperature anisotropies, 
we obtain a B-mode polarization spectrum for three separate cases,
i.e., Cases A, B, and C whose velocity spectra can be seen in
Sec.~\ref{power-summary}.

For Case A, the approximation of the B-mode
polarization spectrum is expressed as
\begin{eqnarray}
\Delta T^{BB} 
&&\approx 
T_0 \left . \sqrt{
{\sqrt{\pi} \over 2 k \sigma} \left(\frac{k L_{\rm mfp}}{3a}\right
)^2  
\left( {\left(\rho_{\rm t}+p_{\rm t}\right) a \over \nu k_{\rm int}} 
{\cal P}_{\rm A}(t_{\rm t-v} ,k_{\rm int}) 
\right)^2
\left({k \over k_{\rm int}}\right) 
}
\right |_{k=l/(\eta_0-\eta_{\rm LSS})}
\nonumber \\
&&
\approx 0.14 \left(\frac{k_{\rm int}}{1.0 {\rm Mpc}^{-1}} \right)^{-7/6} 
 \left( \frac{l}{500} \right)~[\mu {\rm K}],
\qquad l<l_{\rm S},
\label{poral-ani-vis-large-tight}
\end{eqnarray}
\begin{eqnarray}
\Delta T^{BB} 
&&\approx 
T_0 \left . \sqrt{
{
3\sqrt{\pi} \over 2 k \sigma} {1 \over 100} \frac{a}{k L_{\rm mfp}
}  
\left( {\left(\rho_{\rm t}+p_{\rm t}\right) a \over \nu k_{\rm int}} 
{\cal P}_{\rm A}(t_{\rm t-v} ,k_{\rm int}) 
\right)^2
\left({k \over k_{\rm int}}\right) 
}
\right |_{k=l/(\eta_0-\eta_{\rm LSS})}
\nonumber \\
&&
\approx 18.5 \left(\frac{k_{\rm int}}{1.0 {\rm Mpc}^{-1}} \right)^{-8/3} 
 \left( \frac{l}{l_{\rm int}} \right)^{-1/2}~[\mu {\rm K}],
\qquad l_{\rm S}<l<l_{\rm int},
\label{poral-ani-vis-large}
\end{eqnarray}
\begin{eqnarray}
\Delta T^{BB} 
&&\approx 
T_0 \left . \sqrt{
{
3\sqrt{\pi} \over 2 k \sigma} {1 \over 100} \frac{a}{k L_{\rm mfp}
}  
\left( {\left(\rho_{\rm t}+p_{\rm t}\right) a \over \nu k} 
{\cal P}_{\rm A}(t_{\rm t-v} ,k)\right)^2 
} 
\right |_{k=l/(\eta_0-\eta_{\rm LSS})}
\nonumber \\
&&
\approx 18.5 \left(\frac{k_{\rm int}}{1.0 {\rm Mpc}^{-1}} \right)^{-8/3}  
\left( \frac{l}{l_{\rm int}} \right)^{m-2}~[\mu {\rm K}],
\qquad l_{\rm int}<l<l_{\rm f},
\label{poral-ani-vis-small-nodirect}
\end{eqnarray}

\begin{eqnarray}
\Delta T^{BB} 
&&\approx T_0 \left . \sqrt{
{3\sqrt{\pi} \over 2 k \sigma} {1 \over 100} \frac{a}{k L_{\rm mfp}
}
\left( {\left(\rho_{\rm t}+p_{\rm t}\right) k \over \rho_{\rm b} \alpha  a} 
{\cal P}_{\rm A}(t_{\rm t-v} ,k)\right)^2 
} 
\right |_{k=l/(\eta_0-\eta_{\rm LSS})}
\nonumber \\
&&
\approx 5.2 \left(\frac{k_{\rm int}}{1.0 {\rm Mpc}^{-1}} \right)^{-2/3}  
\left( \frac{l}{l_{\rm  f}} \right)^{m}~[\mu {\rm K}],
\qquad l_f<l, 
\label{poral-ani-vis-very-small-nodirect}
\end{eqnarray}
where 
\begin{equation}
l_{\rm S} = k_{\rm S} (\eta_0-\eta_{\rm LSS}) = 1800 .
\end{equation}

For Case B, 
\begin{eqnarray}
\Delta T^{BB} 
&&\approx 
T_0 \left . \sqrt{
\frac{\sqrt{\pi}}{2 k \sigma} 
\left(
\frac{k L_{\rm mfp}}{3a}
\right)^2  
\left( {\left(\rho_{\rm t}+p_{\rm t}\right) a \over \nu k_{\rm int}} 
{\cal P}_{\rm A}(t_{\rm t-v} ,k_{\rm int}) 
\right)^2
\left({k \over k_{\rm int}}\right)  
} 
\right |_{k=l/(\eta_0-\eta_{\rm LSS})}
\nonumber \\
&&
\approx 0.09 \left(\frac{k_{\rm int}}{1.5 {\rm Mpc}^{-1}} \right)^{-7/6} 
 \left( \frac{l}{500} \right)~[\mu {\rm K}],
\qquad l<l_{\rm S},
\label{poral-fs-very-large-tight}
\end{eqnarray}
\begin{eqnarray}
\Delta T^{BB} 
&&\approx 
T_0 \left . \sqrt{
{3\sqrt{\pi} \over 2 k \sigma} {1 \over 100} \frac{a}{k L_{\rm mfp}
} 
\left( {\left(\rho_{\rm t}+p_{\rm t}\right) a \over \nu k_{\rm int}} 
{\cal P}_{\rm A}(t_{\rm t-v} ,k_{\rm int}) 
\right)^2
\left({k \over k_{\rm int}}\right)  
} 
\right |_{k=l/(\eta_0-\eta_{\rm LSS})}
\nonumber \\
&&
\approx 26 \left(\frac{k_{\rm int}}{1.5 {\rm Mpc}^{-1}} \right)^{-8/3} 
 \left( \frac{l}{l_{\rm f}} \right)^{-1/2}~[\mu {\rm K}],
\qquad l_{\rm S}<l<l_{\rm f},
\label{poral-fs-very-large}
\end{eqnarray}
\begin{eqnarray}
\Delta T^{BB} 
&&\approx 
T_0 \left . \sqrt{
{3\sqrt{\pi} \over 2 k \sigma} {1 \over 100} \frac{a}{k L_{\rm mfp}
}
\left( {\left(\rho_{\rm t}+p_{\rm t}\right) k_{\rm int} \over \rho_{\rm b} \alpha a } 
{\cal P}_{\rm A}(t_{\rm t-v} ,k_{\rm int}) 
\right)^2
\left({k \over k_{\rm int}}\right)^{5}  
}
\right |_{k=l/(\eta_0-\eta_{\rm LSS})}
\nonumber \\
&&
\approx 9.9 \left(\frac{k_{\rm int}}{1.5 {\rm Mpc}^{-1}} \right)^{-2/3}  
\left( \frac{l}{l_{\rm int}} \right)^{3/2}~[\mu K],
\qquad l_{\rm f}<l<l_{\rm int},
\label{poral-fs-large}
\end{eqnarray}
\begin{eqnarray}
\Delta T^{BB} 
&&\approx 
T_0 \left. \sqrt{
{3\sqrt{\pi} \over 2 k \sigma} {1 \over 100} \frac{a}{k L_{\rm mfp}
}
\left( {\left(\rho_{\rm t}+p_{\rm t}\right) k \over \rho_{\rm b} \alpha a } 
{\cal P}_{\rm A}(t_{\rm v-f} ,k)\right)^2 
} 
\right |_{l=k(\eta_0-\eta_{\rm LSS})}
\nonumber \\
&&\approx 9.9 \left(\frac{k_{\rm int}}{1.5 {\rm Mpc}^{-1}} \right)^{-2/3} 
 \left( \frac{l}{l_{\rm int}} \right)^{m},
\qquad l_{\rm int}<l.
\label{poral-fs-small-decay}
\end{eqnarray}

Finally, for Case C, the difference with Case B only arises
at the scales smaller than the integral scale as is the case of
temperature anisotropies.
Using Eq. (\ref{power-velo-case-C-blue}), we acquire the B-mode polarization spectrum as
\begin{eqnarray}
\Delta T^{BB} 
&&\approx T_0 \left . \sqrt{
{3\sqrt{\pi} \over 2 k \sigma} {1 \over 100} \frac{a}{k L_{\rm mfp}
}
\left[ \frac{(\rho_{\rm t}+p_{\rm t})k_{\rm int}}{a \alpha \rho_{\rm
      b}} \right]^2 {\cal P}_{\rm A}(k_{\rm int})^2
{\rm e}^{-2\left(k/k_{\rm int}\right)^2}
} 
\right |_{l=k(\eta_0-\eta_{\rm LSS})}
\nonumber \\
&&
\approx 2.4 \left(\frac{k_{\rm int}}{10 {\rm Mpc}^{-1}} \right)^{-2/3}  
{\rm e}^{-\left(l/l_{\rm int}\right)^2}~[\mu {\rm K}],
\qquad l_{\rm int}<l.
\label{poral-fs-small}
\end{eqnarray}

Let us discuss our numerical results of B-mode power spectra in
FIG.~\ref{fig:poral-ani}.
The dashed, solid and dotted lines correspond to Cases A, B, and C.  
It is shown that all lines gradually increase until $l \sim 4000$,
since $\Delta T^{BB} \propto l$ on scales $l<l_{\rm S}$ for all cases. 
Note that we assume the equipartition between magnetic fields and
fluid velocities during the turbulent regime.  On very large scales
(small $l$'s), we may expect violation of this assumption and steeper
slopes for the B-mode power spectra below $l \lesssim 2000$
as is the case with temperature spectra.  
Case A shows continuous declining on $l > 4000$ which is consistent with 
our analytic estimate.  
The increase of $\Delta T^{BB}$ on 
scales $l>l_{\rm f}=17800$ for Cases B and C is caused by the efficient
evolution of the velocity field in the 
free streaming regime due to the lack of dissipation term ${\bf f}$.
For both Cases B and C, $\Delta T^{BB}$ decreases on scales 
$l > l_{\rm int}$.  We can see this decrease in
FIG.~\ref{fig:poral-ani} for Case B whose $l_{\rm int}$ is $22000$.  
For Case C, we cannot find this trend because $l_{\rm int} = 1.4\times
10^5$, which is beyond the range of FIG.~\ref{fig:poral-ani}.  

From the temperature anisotropy spectrum, Case A and a part of Case B are
excluded.   However, the solid line in FIG.\ref{fig:poral-ani}, which overshoots the 
B-mode produced by the gravitational lens effect, is not yet ruled out.
Even the dotted line can provide dominant contribution as 
the B-mode polarization on scales $l> 3000$.

\subsection{Comparison with Linear Calculations}\label{compare:linear}

Here, we compare our results with linear calculations.  Note that 
all previous works investigated the effects of the magnetic fields on 
CMB anisotropies and polarization are based on linear perturbation theory.  

In the Appendix, we summarize the linear perturbation analysis and the
approximated estimations of temperature anisotropy and 
polarization spectra following the analysis by Subramanian and Barrow
(\cite{s-b-temp,s-b-poral}).

In FIG.~\ref{fig:temp-ani}, we plot the temperature power spectrum of
the linear calculation with $B=28 \rm nGauss$ at $k=1.5 \rm Mpc^{-1}$ and
$n=4$ as the gray line for comparison.  It is shown that 
differences with the nonlinear calculation (the solid line) appear on
scales $l < 2000$, and $l> 18000$.

On large scales, we can ignore the diffusion in the early epoch.  In
the nonlinear calculation, $v$ immediately approaches the Alfv\'en
velocity $v=v_{\rm A}$ due to the nonlinear coupling during the turbulent
regime {if there exist strong magnetic fields in the very early universe}.  
On the other hand, the growth of the velocity induced by the
magnetic pressure gradient in the linear calculation turns out to be
much slower as is shown in the Appendix (Eq. (\ref{v-nonvis})).  Accordingly we
find {a smaller} temperature power spectrum (and polarization) for
the linear calculation.  {Note that even for the nonlinear calculation, 
there may be the case in which the equipartition could not be achieved
by the end of the turbulent regime if the primordial magnetic fields
on the large scale were not strong enough.  In this case, we rather
expect to have similar behaviors of the power spectrum to the linear
one on large scales (small $l$'s)}.  

Once the viscosity becomes efficient, the velocity follows the linear
solution even in the case of the nonlinear calculation (see
Eqs.~(\ref{viscous-velo}) and (\ref{v-vis})).  Unlike the perturbations without the
magnetic fields which suffer severe damping, the velocity (Alfv\'en
mode) induced by the magnetic fields can survive within the Silk
scale for both nonlinear and linear calculations.  
The amplitude of the velocity is determined by the Alfv\'en
velocity, i.e., the amplitude of the magnetic fields.  Therefore both
linear and nonlinear calculations give almost identical results, which
are $\Delta T^{TT}(l) \propto l^0$, between $2000 (\sim l_{\rm S}) < l <
18000$.  

On small scales, $l > 18000$, the linear calculation shows steeper
rise of the power spectra for both Cases A and B than the nonlinear
calculation as shown in FIG. 4.  There also exists exponential
damping in the linear calculation on smaller scales.  On the other
hand, nonlinear calculation shows less rise and no damping in the
power spectra.  

The reason why the power spectra of the nonlinear calculation have
less prominent peaks on the small scales (large $l$'s) is due to the
cascade decay of the magnetic fields during the turbulent regime.  As
shown in FIG. 2, the peak of the magnetic field power spectrum shifts
to larger scales.  The condition of this shift is determined by
$t_{\rm eddy}\equiv L_{\rm int}/v_{\rm A} =1/H$ as is discussed in
Sec.~\ref{velocity_turb}.  Then the cutoff scale (or integral scale)
has been frozen since the transition from the turbulent regime to the
viscus regime.  Accordingly the comoving wave number of the cutoff can
be written and $k_{\rm int} = aH/v_{\rm A}|_{z_{t-v}}$ as is obtained
in Eq.~(\ref{eq:int_viscus}).  On the other hand, the cutoff scale of
the linear calculation is always determined by the diffusion condition
$k_{\rm c}=k_{\rm S}/v_{\rm A}$ (see Appendix).  The power spectrum
increases toward small scales as long as $k<k_{\rm c}$.  Then it
starts to show exponential damping at $k_{\rm c}$.  In Cases A and B,
$k_{\rm int} < k_{\rm c}$ at the recombination epoch.  Therefore
nonlinear calculation shows less peaks in the power spectra for these
cases.  The diffusion damping scale is determined by $k_{\rm c}=k_{\rm
S}/v_{\rm A} \propto 1/B$ for both linear and nonlinear calculations.
Therefore one might expect to have same damping behaviors for both
linear and nonlinear calculations.  However, since the magnetic field
strength of the nonlinear calculation on the scales smaller than
$1/k_{\rm int}$ is smaller than the one of the linear calculation as
well as temperature anisotropies, we expect to have the smaller
diffusion damping scale for the nonlinear calculation than the linear
calculation, i.e., $k_{\rm c}^{\rm nonlinear} > k_{\rm c}^{\rm
linear}$. Within the range of FIG.~4, we can only see the diffusion
(exponential) damping of the linear calculation. The nonlinear
calculations for Cases A and B only show much milder damping due to
the cascade decay during the turbulent regime below the integral
scale.

In Case C, $k_{\rm int} > k_{\rm c}$ at the recombination epoch.
Therefore we ought to obtain very similar temperature power spectra
for both linear and nonlinear calculations.

\section{Conclusion}
In this paper we study the effect of the primordial magnetic fields on
the CMB temperature and polarization anisotropies.  In particular the
nonlinear evolution of the magnetic fields and the resulting Alfv\'en
modes of the fluid velocities, which are the source of
temperature anisotropies and polarization, are appropriately included
based on the cosmological MHD simulation by Banerjee and Jedamzik
\cite{b-j-letter,b-j-paper}.  Diffusion and direct cascade processes
are properly taken into account.

We separate the evolution into three regimes, i.e., turbulent,
diffusion, and free streaming.  In the turbulent regime, the advective
term, which is essentially nonlinear, provides the dominant
contribution on the evolution of Alfv\'en modes.  Viscosity plays an
important role in the viscous regime, while the drag term takes over
the task of the viscous term in the free streaming regime.

By combining the numerical simulations of three regimes, we obtain a
comprehensive evolution history of the magnetic fields and Alfv\'en
modes.  We find the relation between the integral (or coherent) scale
on which the magnetic energy peaks, and the maximum
magnetic field strength, or equivalently, the magnetic field strength
at the integral scale as shown in FIG. \ref{fig:magallowed}.

We divide the evolution of the magnetic fields and the Alfv\'en
modes into three cases, i.e., Cases A, B, and C.  For Case A, the
integral scale is still in the viscous regime at the recombination
epoch, while the integral scale is in the free streaming regime for
Cases B and C.  The direct cascade process takes place during the 
turbulent regime for all cases. In the free streaming regime the
direct cascade process as well as the diffusion process works for 
Case C at the integral scale.  
The resultant velocity spectra of Alfv\'en modes are quite
different between three cases.

From the velocity spectra, we calculate CMB temperature anisotropy and
(B-mode) polarization spectra.  We make comparisons between nonlinear and linear
calculations and find differences on both large and small scales. On
large scales $l< 2000$, both CMB anisotropy and polarization spectra
have flat and blue spectra for the nonlinear and linear calculations,
respectively.  This difference is caused by the inclusion of the
advective term for the nonlinear calculations. This difference gives
stronger constraints for the nonlinear case in the intermediate
angular scale using CMB observations such as WMAP and BOOMERANG. Note
that the possible non-equipartition in the turbulent regime may make
the difference between nonlinear and linear calculations small.

Using WMAP and BOOMERANG, we set rough upper limits for the comoving
magnetic field strength as $B < 28 \rm nGauss$ and the comoving integral
scale as $k_{\rm int} > 1.5 \rm Mpc^{-1}$ if the equipartition 
is valid in the turbulent regime. If this assumption is violated, we
still can set rough upper limits as $B < 30 \rm nGauss$ and 
$k_{\rm int} > 1.2 \rm Mpc^{-1}$.

For the polarization spectra, we also expect higher signals on large
scales by nonlinear calculations than linear ones.  The signal may
exceed the B-mode polarization from the gravitational lenses if
primordial magnetic fields exist.

On small scales, nonlinear calculations show milder damping of
temperature anisotropies and polarization than linear calculations in
Cases A and B.  We expect to have both temperature anisotropy and
polarization spectra even beyond $l >10000$.  The peak values of temperature
anisotropy and B-mode polarization spectra are approximately $40\mu
\rm K$ and a few $\mu \rm K$ depending on the peak scale $l_{\rm int}$
(or the integral scale $k_{\rm int}$).

Note that we consider only the case with the spectral index of the magnetic
field spectrum $n>3/2$ in this paper, while
the extension to the case with $0<n<3/2$ is straightforward.  

Various observation projects for the small scale temperature anisotropies
and polarization are planed \cite{obs-project}.    
These observations may find the evidence of primordial magnetic
fields or at least will set stringent constraints.  
Perhaps the best constraint will be provided by the really fine scale
data, i.e., $l > 10000$.

Finally, another and possibly stronger constraint on the field
strength of putative primordial magnetic fields might be derived from
the deflection of ultra high energy cosmic rays (UHECR).  
Dolag et al.~\cite{Dolag05} simulated the evolution of magnetic fields
using a MHD SPH code and solved the propagation of UHECR.  Their seed
magnetic fields were only $2 \times 10^{-12}$Gauss and they found that
strong deflections of UHECR with arrival energies $E=4\times 10^{19}$
within a distance of $100$Mpc occur only when UHECR cross galaxy
clusters.  Therefore the arrival directions of UHECR mostly trace the
source positions on the sky.  On the other hand, if there exists the
magnetic fields of $10$nGauss as a seed, we expect to have very strong
deflections for UHECR with arrival energies even above $10^{20}$eV.
In near future, a huge amount of UHECR will be observed by the air
shower experiments such as the Pierre Auger observatory \cite{auger1}
whose angular resolution is $\sim 0.6$ degree \cite{auger2}.  If the
observed UHECR with $10^{20}$eV, which only can arrive from nearby
sources ($\lesssim 100$Mpc) due to GZK cutoff \cite{gzk}, do trace the
large scale structure, we will be able to set a very stringent upper
limit for the primordial magnetic field strength.  
Note that the propagation and the deflection of UHECR are still open
questions.  Contrary to Dolag et al., Sigl et al. found larger
deflection of UHECR in their simulation \cite{sigl}.  They claimed that the
sources are strongly magnetized ($\sim \mu$Gauss) and the deflection
angle can be of order $20^\circ$ up to $10^{20}$eV even if the
extragalactic magnetic fields of the observer are negligible ( $\ll
0.1\mu$Gauss).  If this is the case, however, we can still set an
upper limit to the magnetic filed strength, if the observed UHECR
above $10^{20}$eV do trace the large scale structure.
In near future, a
huge amount of UHECR will be observed by the air shower experiments
such as the Pierre Auger observatory \cite{auger1} whose angular
resolution is $\sim 0.6$ degree \cite{auger2}.  If the observed UHECR
with $10^{20}$eV, which only can arrive from nearby sources ($\lesssim
100$Mpc) due to GZK cutoff \cite{gzk}, do trace the large scale
structure, we will be able to set a very stringent upper limit for the
primordial magnetic field strength.

\acknowledgements

We would like to thank Karsten Jedamzik for many crucial comments.
One of the author (N.S.) would like to thank Wayne Hu for useful
comments and also thank for kind hospitality of Kavli institute for
Cosmological Physics at the University of Chicago. 
N.S. is supported by a Grant-in-Aid for Scientific Research from the
Japanese Ministry of Education (No. 17540276).

\appendix
\section{linear perturbation}

In this appendix, we review the linear perturbation analysis of the magnetic fields
in the early universe.
For a detailed discussion,
see the references \cite{s-b-temp, m-k-k, s-b-poral}.

We assume that
the metric perturbations of the vector mode are expressed as Eq. (\ref{metric}).
The stress energy tensors are
divided into the fluid-part, $T_{\rm F}^{\mu \nu }$, and the magnetic field part, $T_{\rm B}^{\mu \nu }$.
In the fluid part, we take into account of the viscosity between
photons and baryons (electrons) 
\begin{equation}
T_{\rm F}^{\mu \nu }=(p_{\rm t}+\rho_{\rm t} )U^\mu U^\nu +p_{\rm t} g^{\mu \nu } 
-\eta H^{\mu \alpha }H^{\nu \beta }W_{\alpha \beta },
\end{equation}
\begin{equation}
H^{\alpha \beta }\equiv g^{\alpha \beta }+U^\alpha U^\beta,~~~ 
W_{\alpha,\beta }\equiv U_{\alpha ;\beta }+U_{\beta ;\alpha }-{\frac 23}g_{\alpha
\beta }U^\gamma {}_{;\gamma } ,~~~
\eta ={\frac 4{15}} \rho _\gamma L_{\rm mfp}. 
\end{equation}
Here $U^\mu $ is the four-velocity and is written as 
\begin{equation}
U^0=1/a,~~~~U^i=u^i/a,
\end{equation}
where $u^i$ are velocities.  
Now we are interested in the vector mode of the stress energy tensor perturbations, 
so that the velocity filed 
is divergence free and is decomposed as Eq. (\ref{vector-deco}).

The magnetic part of the stress energy tensor is expressed as
\begin{equation}
T _ {{\rm B}  ij}({\bf k}) =  \frac{1}{4 \pi} 
\int d^3 p \left[B_{i}({\bf p}) B_j({\bf k-p})
-\frac{1}{2} \delta_{ij} B_l ({\bf p})B_l({\bf k-p})
\right].
\label{mag-ene-str}
\end{equation}
For obtaining the vector mode of the magnetic part, 
we introduce a projection tensor  onto the transverse plane as 
\begin{equation}
P_{ij}({\bf k}) \equiv \delta_{ij} - k_{ i} k_j / k^2.
\label{tangent}
\end{equation}
Then the vector mode of Eq. (\ref{mag-ene-str}) can be written as 
\begin{equation}
T^V _{{\rm B}ij} = (P_{in} k_j +P_{jn} k_i) k_m T_{Bmn}/k^2.
\end{equation}
From the conservation law of the energy momentum tensors $T_F$ and $T_B$ , 
we obtain the Euler equation 
\begin{equation}
(\dot u_i-\dot V_i)
+ (1-3 c_{\rm st}) \frac{\dot a}{a} (u_i-V_i) +\frac{\nu}{a} \frac{k^2}{\rho_{\rm t}+ p_{\rm t}} (u_i -V_i)
=\frac{k \Pi _i}{a^4 (\rho_{\rm t}+ p_{\rm t})},
\label{line-eular}
\end{equation}
where dots represent the derivatives with respect to the conformal time and the comoving pressure gradient term of the magnetic field
$\Pi_i$ is defined as 
\begin{equation}
{\Pi _i \over a^4}=P_{in}  {\hat k}_m T_ {\rm B}^{mn}.
\end{equation}

As discussed in Sec. \ref{seccalcmb},
we use  $v_i$ which is defined as $v_i\equiv u_i-V_i$.
We can rewrite Eq.~(\ref{line-eular}) in terms of $v_i$.
When we evaluate Eq. (\ref{line-eular}) at large scales
where the viscosity can be neglected as is the case of the turbulent
regime in the nonlinear calculations, 
we obtain as the following approximation in the radiation dominated epoch
\begin{equation}
v_i =\frac{3  k \Pi _i \eta}{4 (R+1) \rho _{\gamma }a^4 },
\label{v-nonvis}
\end{equation}
where $R =3\rho _{\rm b} /4 \rho_\gamma $.
We refer the scales where Eq. (\ref{v-nonvis}) is valid
as ``no diffusion scales''.

At small scales where 
the viscosity is dominant over the cosmological expansion, 
which we call as ``viscous scales'', 
we can neglect the expansion term in Eq. (\ref{line-eular})
and obtain 
\begin{equation}
v_i =\frac{15 \Pi _i}{4 \rho _{\gamma }  a^3 k  L_{\rm mfp}}, 
\label{v-vis}
\end{equation}
with  employing the terminal velocity approximation.  


Matching Eqs. (\ref{v-nonvis}) and (\ref{v-vis}), we acquire 
the transition scale between the no diffusion scales 
and the viscous scales as 
\begin{equation}
k_{\rm v} \sim \left[ 
{5 a (1 
+R) \over \eta  L_{\rm mfp} }\right]^{1/2}.
\end{equation}

At the smaller scales than the photon mean free path, which we refer as 
``free streaming scales'', baryons and photons are decoupled and 
we can no longer adopt the diffusion approximation, Eq.~(\ref{line-eular}).
Instead, we must introduce the drag force term 
${\dot \tau}v_{ i}/{R}$ in the baryon Euler equation as 
\begin{equation}
\dot v_{i} +\frac{\dot a}{a} v_{i}
+\frac{\dot \tau}{R}
v_{ i}=\frac{k \Pi_i}{\rho_{\rm b} a^4}.
\end{equation}
When we neglect the cosmological expansion term and 
apply the terminal velocity approximation we obtain
\begin{equation}
v_i=\frac{3 k \Pi _i L_{\rm mfp}}{4 a^5 \rho_{\gamma }}.
\label{v-free}
\end{equation}
The transition scale 
between the viscous scales and the free streaming scales 
is $k_{\rm f}$ as discussed in Sec. \ref{velo-viscoussection}
because Eqs.~(\ref{v-vis}) and (\ref{v-free}) are identical 
to Eqs.~(\ref{viscous-velo}) and (\ref{fs-velo}).

For the comparison with the nonlinear results
we calculate the power spectrum of the velocity fields.
From the above results 
$v_i$ is rewritten as
\begin{equation}
v_i =kX(\eta)  \Pi_i,
\end{equation}
\begin{equation}
X(\eta)=
\cases{
3 \tau/ 4 \rho_{\gamma } a^4(1+R),& \quad $k< k_{\rm v} $,\cr
15 /4 \rho_{\gamma }a^3 k^2  L_{\rm mfp},& \quad $k_{\rm v}<k<k_{\rm f}$, \cr
3 L_{\rm mfp}/4 \rho_{\gamma } a^5,& \quad $k_{\rm f} <k .$ \cr
}
\end{equation}
The ensemble average of the velocity fields leads to 
\begin{equation}
\langle |v_i |^2 \rangle =k^2 X(\eta)^2 \langle |
\Pi_i |^2 \rangle.
\end{equation}
By following the procedure of ref. \cite{m-k-k},
$\left\langle | \Pi_i |^2\right\rangle$ can be solved as 
\begin{equation}
\langle|
 \Pi _i
|^2\rangle =\frac{1}{(4 \pi)^2}\int_{0}^{\infty }{dq^3} {{\cal P}_B(q)\over q^3} 
{{\cal P}_B(|({\bf k}+{\bf q})|) \over |({\bf k}+{\bf q})|^3 } 
(1-\mu ^{2})\left[ 1+{\frac{(k+2q\mu )(k+q\mu )}{
(k^{2}+q^{2}+2kq\mu )}}\right] ,
\end{equation}
where $q=|{\bf q} |$ and $\mu={\bf k} \cdot {\bf q}/(qk)$.  Now, we
assume that the magnetic power spectrum is the power law spectrum with
the spectral index $n$ as 
\begin{equation}
{\cal P}_B={k^3 \over 2\pi^2} \langle|B_{\rm
  comov}|^2 \rangle=B_{-9} ^2 \left({k \over k_{\rm n} }\right)^n, 
\end{equation}
where the strength of the  magnetic fields is $B_{-9}$nGauss at the wave length $k_{\rm n}$.

Assuming the power law spectrum for the magnetic fields, we can obtain the analytic
approximation of $\langle | \Pi_i |^2 \rangle$ 
\cite{s-b-temp,m-k-k, s-b-poral}.  Accordingly,  the
power spectrum of the fluid velocity leads to 
\begin{equation}
{\cal P}_{v}(k)={k^3 \over 2 \pi^2} \langle | v_i |^2 \rangle 
\approx k^2 B_{-9} ^2 X(\eta)^2 I(k)^2 ,
\end{equation}
where the mode coupling $I(k)$ is different for the spectrum with the
index $n<3/2$ and $n>3/2$.  
From the reference
\cite{s-b-temp},in the case of $n<3/2$, the mode coupling $I(k)$ is
approximated as
\begin{equation}
I^2(k) = {1 \over 32 \pi^4}{2 \over 3 n}  
{\left( k\over k_{\rm n} \right)^{2n}}.
\end{equation}
In the case of $n>3/2$, 
\begin{equation}
I^2(k) = {1 \over 32 \pi^4}{7 \over 15 (2n-3)} 
\left({ k\over k_{\rm n}} \right)^{3} \left({k_{\rm c} \over k_{\rm n} } \right)^{2n-3},
\label{bluemodecoupling}
\end{equation}
where $k_{\rm c}$ is the cutoff scale defined as $k_{\rm c} \equiv k_{\rm S}/v_{\rm A}$ \cite{j-k-o,s-b-nonlinear}.  
This cutoff of the power spectrum is caused by the dissipation due to
the drag force.

For the comparison with the nonlinear results in this paper, 
we focus on the case with the blue spectrum $n>3/2$.  
Using Eq. (\ref{bluemodecoupling}), we obtain the power spectrum of the fluid velocity
as 
\begin{equation}
{\cal P}_{v}(k) \approx k^2 \left[{3 \eta \over 4 \rho_{\gamma
} a^4 (1+R)} \right]^2 I(k)^2 \propto k^5, \qquad k<k_{\rm v},
\label{power-line-nodiffusion}
\end{equation}
\begin{equation}
{\cal P}_{v}(k) \approx  k^2 \left[{
15  \over 4 \rho_{\gamma }a^3 k^2 L_{\rm mfp}} \right]^2 I(k)^2
\propto k, \qquad k_{\rm v}<k<k_{\rm f},
\label{power-line-vis}
\end{equation}
\begin{equation}
{\cal P}_{v}(k) \approx  k^2 
\left[{3 L_{\rm mfp} \over 4 \rho_{\gamma } a^5
} \right]^2 I(k)^2
\propto k^5, \qquad k_{\rm f}<k.
\label{power-line-fs}
\end{equation}

From the velocity power spectrum, CMB temperature anisotropy and polarization
spectra can be calculated.  Subramanian and Barrow
\cite{s-b-temp, s-b-poral} obtained the following results with using the
small angle approximation.  

In the case of $k \sigma \ll 1$,
\begin{equation}
\Delta T^{TT} =T_0 
\sqrt{{l (l+1) C(l) \over 2 \pi}}  
\approx T_0 \left.  \sqrt{{\pi \over 2} {\cal P}_v (k) }
\right |_{k=l/(\eta_0-\eta_{\rm LSS})},
\label{line-temp-ani-large}
\end{equation}
\begin{equation}
\Delta T^{BB} =
T_0 \sqrt{{l (l+1) C(l) \over 2 \pi}} 
\approx \left. T_0
\sqrt{ {\pi \over 2}  \left( {k L_{\rm mfp} \over 3 a} \right)^2 P_v(k )} 
\right|_{k=l/(\eta_0-\eta_{\rm LSS})}.
\label{line-poral-ani-large}
\end{equation}
And for the case of $k \sigma \gg 1$,
\begin{equation}
\Delta T^{TT}=T_0 \sqrt{{l (l+1) C(l) \over 2 \pi}}
 \approx T_0 \left.  \sqrt{{\pi^{1/2} \over2 k \sigma}  P_{v}(k )} 
 \right |_{k=l/(\eta_0-\eta_{\rm LSS})},
\label{line-temp-ani-small}
\end{equation}
\begin{equation}
\Delta T^{BB} =T_0 \sqrt{{l (l+1) C(l) \over 2 \pi} }
\approx T_0 \left . \sqrt{{\pi^{1/2} \over 2 k \sigma}  \left(\frac{k L_{\rm mfp}}{3 a}\right)^2  P_{v}(k )} 
\right |_{k=l/(\eta_0-\eta_{\rm LSS})}.
\label{line-poral-ani-small}
\end{equation}

The temperature anisotropy spectrum induced by the magnetic fields with
spectral index $n>3/2$ in the low multipoles is given by substituting
Eq. (\ref{power-line-nodiffusion}) into
Eq. (\ref{line-temp-ani-large}) and the one in the high multipoles is
given by Eqs. (\ref{power-line-vis}) and (\ref{power-line-fs}) into
Eq. (\ref{line-temp-ani-small}) as 
\begin{equation}
\Delta T^{TT}
\approx 
5.4 \ B_{-9} ^2 \left(l \over 1000 \right) I (k_l) 
~[\mu {\rm K}] \propto l^{5/2},
\ 
l<l_{\rm v},
\end{equation}
\begin{equation}
\Delta T^{TT}
\approx 
13.0 \ B_{-9} ^2 \left(l \over 2000 \right)^{-3/2} I (k_l)
~[\mu {\rm K}] \propto k^0,
\
l_{\rm v}<l<l_{\rm f},
\end{equation}
\begin{equation}
\Delta T^{TT}
\approx 
0.4 \ B_{-9} ^2 \left(l \over 20000 \right)^{1/2} I (k_l) 
~[\mu {\rm K}] \propto l^2,
\
l_{\rm f}<l<l_{\rm c},
\end{equation}
\begin{equation}
I(k_l)=I (k) |_{k=l/(\eta_0 -\eta_{\rm LSS})} ,
\end{equation}
where $l_{\rm v}=k_{\rm v} (\eta_0 -\eta_{\rm LSS})$ 
and $l_{\rm c}=k_{\rm c} (\eta_0 -\eta_{\rm LSS})$.

The B-mode polarization spectrum induced by the magnetic fields with spectral
index $n>3/2$ is acquired by substituting
Eq. (\ref{power-line-nodiffusion}) into
Eq. (\ref{line-poral-ani-large}) and substituting
Eq. (\ref{power-line-vis}) into Eq. (\ref{line-poral-ani-small}) as 
\begin{equation}
\Delta T^{BB}
\approx 
0.04 \ B_{-9} ^2 \left(l \over 1000 \right)^{2} I (k_l) 
~[\mu {\rm K}] \propto l^{7/2},
\ 
l<l_{\rm v},
\end{equation}
\begin{equation}
\Delta T^{BB}
\approx 
0.12  \ B_{-9} ^2 \left(l \over 2000 \right)^{-1/2} I (k_l) 
~[\mu {\rm K}] \propto l,
\
l_{\rm v}<l<l_{\rm f}.
\end{equation}

\end{document}